\documentclass[iop,apj,twocolumn,numberedappendix,appendixfloats]{openjournal}
\usepackage{fix-cm}
\usepackage{lastpage}
\usepackage{graphicx}
\graphicspath{ {./figures/} }
\usepackage{xcolor}

\usepackage{newtxtext,newtxmath,enumitem,multirow,ctable,nccmath,siunitx,tensor,amsmath}
\usepackage{bm}

\usepackage{bbold}
\usepackage{savesym}
\usepackage{hyperref}
\savesymbol{tablenum}
\hypersetup{breaklinks,colorlinks,citecolor=blue,urlcolor=cyan}

\newcommand{\dd}{\mathrm{d}}
\newcommand{\bvec}[1]{\boldsymbol{\mathbf{#1}}}
\renewcommand{\pi}{\uppi}

\DeclareSIUnit\parsec{pc}
\DeclareSIUnit\sqd{deg^2}
\DeclareSIUnit\arcmin{arcmin}

\shorttitle{Exact likelihood for correlation functions}
\shortauthors{V. Oehl \& T. Tröster}

\begin{document}
\title{The exact non-Gaussian weak lensing likelihood: A framework to calculate analytic likelihoods for correlation functions on masked Gaussian random fields}
\author{Veronika Oehl$^{\star}$}
\author{Tilman Tröster}
\affiliation{Institute for Particle Physics and Astrophysics, ETH Zurich, 8093 Zurich, Switzerland}
\thanks{$^\star$E-mail: \href{mailto:veoehl@phys.ethz.ch}{veoehl@phys.ethz.ch}}
\journalinfo{The Open Journal of Astrophysics}

\begin{abstract}
We present exact non-Gaussian joint likelihoods for auto- and cross-correlation functions on arbitrarily masked spherical Gaussian random fields. 
Our considerations apply to spin-$0$ as well as spin-$2$ fields but are demonstrated here for the spin-$2$ weak-lensing correlation function.

We motivate that this likelihood cannot be Gaussian and show how it can nevertheless be calculated exactly for any mask geometry and on a curved sky, as well as jointly for different angular-separation bins and redshift-bin combinations. 
Splitting our calculation into a large- and small-scale part, we apply a computationally efficient approximation for the small scales that does not alter the overall non-Gaussian likelihood shape. 

To compare our exact likelihoods to correlation-function sampling distributions, we simulated a large number of weak-lensing maps, including shape noise, and find excellent agreement for one-dimensional as well as two-dimensional distributions. 
Furthermore, we compare the exact likelihood to the widely employed Gaussian likelihood and find significant levels of skewness at angular separations $\gtrsim \ang{1}$, such that the mode of the exact distributions is shifted away from the mean towards lower values of the correlation function. 
We find that the assumption of a Gaussian random field for the weak-lensing field is well valid at these angular separations. 

Considering the skewness of the non-Gaussian likelihood, we evaluate its impact on the posterior constraints on $S_8$.
On a simplified weak-lensing-survey setup with an area of $\SI{10000}{\sqd}$, we find that the posterior mean of $S_8$ is around $2.5\%$ higher when using the non-Gaussian likelihood, a shift comparable to the precision of current stage-III surveys. 
\end{abstract}
\maketitle
\section{Introduction}

Spherical two-dimensional fields as projections of the three-dimensional structure surrounding us are ubiquitous in cosmology and provide a range of observables of the large-scale structure of the Universe.
The most prominent and early-time example is the cosmic microwave background (CMB) with its temperature and polarization fields.
Late-time examples include galaxy number counts and cosmic shear; the latter will be the example of choice for this work.

Primordial density fluctuations very closely follow a Gaussian distribution. 
On large scales, gravitational collapse remains linear, and since projection along the line of sight is a linear operation, the resulting two-dimensional mass and weak lensing maps can be assumed to be Gaussian on sufficiently large scales.

Gaussian random fields are completely characterized by their mean and two-point function.
The latter can be expressed in configuration (correlation function) and harmonic space (power spectrum).

As they are mathematically equivalent for continuous fields, both representations of the two-point function can be predicted from the cosmological standard model, requiring only a small set of parameters.
Conversely, cosmological parameters can be constrained from measurements of two-point summary statistics, which has been the standard approach to parameter inference from cosmological fields for past and current surveys.
Whenever such a parameter inference is attempted, a likelihood $p(\bvec{x} | \bvec{\vartheta})$ encapsulating the probability to obtain the data $\bvec{x}$ given a set of model parameters $\bvec{\vartheta}$ needs to be specified.

In our case, the data $\bvec{x}$ are a set of measured two-point functions, which can be obtained on different angular scales, from different tomographic bins, or even different probes.
In most applications, these observables are not independent because they are sourced by the same matter density field. As a result, the joint $n$-dimensional likelihood function must be considered.

Two-point functions generally do not follow a Gaussian distribution even if the fields they are measured on are perfectly Gaussian, as we will show in Sect.~\ref{sec:ana}.
However, likelihoods are often assumed to be Gaussian due to the comparably simple analytic form and because it is a conservative choice if the exact form is unknown. 

This has also been the case for two-point functions in cosmology where the use of a Gaussian likelihood can be justified by the central limit theorem: an estimator for a two-point summary statistic will assume a Gaussian likelihood if it involves a sum over a sufficiently large number of random variables. 
Considering that weak-lensing observations are currently limited to a fraction of the celestial sphere mostly due to survey constraints and bright foregrounds, which restricts the analyses to small angular separations, the assumption of a Gaussian likelihood is an excellent approximation \citep{joachimi2021a,friedrich2021a,hall2022a} and has been used for stage-III cosmic-shear analyses such as the Subaru Hyper Suprime-Cam \citep[HSC,][]{aihara2018a,dalal2023a,li2023a} survey, the Dark Energy Survey \citep[DES,][]{collaboration:2016a,secco2022a,amon2022a}, the Kilo-Degree Survey \citep[KiDS,][]{kuijken2019a,asgari2021a,li2023b}, and also comparisons amongst them \citep[e.g.][]{dark-energy-survey-and-kilo-degree-survey-collaboration2023a}.

On large angular scales, the number of available modes is limited, leading to significant cosmic variance. In this regime the central limit theorem no longer holds and the non-Gaussianity of the likelihood becomes important \citep{joachimi2021a}.
Meanwhile, the Gaussianity of the underlying field is well motivated on these largest scales \citep{clerkin2016a}. 
The present work addresses the likelihood on these large, Gaussian scales.
On smaller scales, non-linear structure formation and baryonic effects cause the field to deviate from Gaussianity.
\citet{hall2022a} extend the Wishart distribution by introducing small deviations from Gaussianity at the field level and find that, on these small scales, a Gaussian likelihood remains adequate -- provided the covariance is corrected to include the trispectrum contribution.

Discrepancies between inferences of the same cosmological parameter from early- and late-time observables emerged with increasing statistical power and measurement precision over the past ten years. 
For weak lensing, a parameter that can be constrained well is the lensing amplitude or structure-growth parameter $S_8$.
When inferred from the primary CMB, this parameter is generally found to be higher than when inferred using low-redshift probes such as weak lensing \citep[][]{planck-collaboration2020a,amon2022a,li2023a,li2023b}.
Even though the tension in $S_8$ between late and early Universe probes seems to be somewhat alleviated by recent combined analyses \citep[][]{dark-energy-survey-and-kilo-degree-survey-collaboration2023a}, it is not widely considered as solved.

Such discrepancies can arise if the chosen cosmological model does not describe our Universe consistently from early to late times or more generally if the parametric model is wrong.
Systematic effects in measurement and statistical inference that have not been fully accounted for can also introduce biases.
The likelihood function plays a crucial role in the inference process. Therefore, errors in its formulation could introduce inaccurate information about the data and directly influence posterior constraints. 

The possible importance of deviations of the likelihood from Gaussianity on large scales has long been acknowledged in the CMB community.
For instance, \citet{upham2020} derive the exact likelihood for the pseudo-power spectrum and show that it is highly non-Gaussian on large scales.
In the latest CMB analyses, hybrid methods are employed with a pixel-based likelihood for the large scales and a Gaussian approximation for the small scales \citep{planck-collaboration2020b}.

In the weak-lensing community, the likelihood of the two-point summary statistic has gained attention as well \citep{sellentin2017,sellentin2018,10.1093/mnras/stz558,lin2020a,upham2021a,hall2022a}, but so far, the real-space correlation-function likelihood has not been considered analytically. Additionally, the impact on posterior constraints for cosmological parameters has not been explored nor have beyond-Gaussian likelihoods been explicitly applied to weak-lensing data.

\citet{lin2020a} infer a subset of the parameters from weak-lensing mock observations using beyond-Gaussian approximations to the exact correlation-function likelihood within the flat-sky approximation. 
They report no significant impact on the posterior constraints. 
However, the approximations to the likelihood rely on a large number of mock observations that might not always be available or feasible to set up for higher-dimensional parameter spaces.
As the number of simulations is finite, low 
-probability tails might not be represented well.
Furthermore, large angular scales are explicitly excluded in this study but are the regime where the likelihood becomes increasingly non-Gaussian.

Large angular scales will also be accessed by upcoming stage-IV weak-lensing surveys, such as the Legacy Survey of Space and Time (LSST) at the Vera C. Rubin Observatory \citep{collaboration2009a} or Euclid \citep{laureijs2011a}.
They cover much larger fractions of the sky and therefore technically allow the measurement of two-point functions at larger angular separations, such that the Gaussian approximation to the two-point-function likelihood might no longer be sufficient.

As the configuration-space correlation function is still routinely used in weak-lensing analyses, one of the reasons being that galaxies as the tracers of the cosmic-shear signal do not naturally form a continuous field on the sky, we will investigate the exact likelihood of two-point functions on the sphere with a focus on the weak-lensing correlation function.

While hierarchical field-level approaches offer a principled alternative to summary-statistic inference \citep{alsing2016a, loureiro2023a}, the application to the data volumes and sky coverage of upcoming weak-lensing surveys remains challenging due to the high computational demand of these methods.

Parameters can also be inferred by comparing the data to a large set of simulations. 
This approach, known as simulation-based inference \citep{cranmer2020a}, bypasses the need to explicitly specify likelihoods \citep[for weak lensing applications see e.g.][]{taylor2019a,wietersheim-kramsta2024a,jeffrey2024a}. 
However, simulation-based inference is limited to the coverage of the simulated parameter space, comes with substantial computational cost, and often lacks the convergence guarantees of traditional methods. 
Accurate likelihoods for summary statistics like the correlation function therefore remain an essential and complementary tool to these new alternative approaches.

In this work, we show that the correlation-function likelihood $p(\bvec{\hat\xi} | \bvec{\vartheta})$ can be calculated analytically. 
Here, $\bvec{\hat{\xi}}$ denotes a vector of measurements of the correlation function obtained from the random field through an estimator (denoted by the hat).
We investigate the exact likelihood particularly for the large scales where, for the cosmological fields in question, the assumption of a Gaussian field holds.
  
The analytical likelihood can be used to quantify its deviation from Gaussianity.
In a further step, this allows predicting the impact on the posteriors and therefore control a possible systematic causing the $S_8$ tension.

The paper is organized as follows: first, in Sect.~\ref{sec:ana}, we will show that the correlation-function likelihood is generally not Gaussian. 
We will then introduce the formalism that allows the calculation of the likelihood of any two-point function on a masked Gaussian random field.  
Throughout the work, we will focus on the application to the weak-lensing correlation function.
Due to computational constraints, this exact calculation can only be applied to the largest contributing scales.
In the next section, Sect.~\ref{sec:full}, we will therefore extend the analytic likelihood to the full correlation function by means of an approximate small-scale part.
In Sect.~\ref{sec:wlapp}, we introduce simulated sampling distributions of the correlation function, which we then compare to our exact likelihood in Sect.~\ref{sec:res}.
There, we also compare our exact likelihoods to the usually employed Gaussian likelihoods and show first results of the effects on the obtained posterior constraints for $S_8$.
We close with an outlook onto planned improvements and possible applications.  
In the appendix, we disclose more technical details; in particular, we discuss and justify the assumption of Gaussianity of the weak-lensing fields in the context of this work.

\section{Analytic Multidimensional Likelihoods of Two-Point Functions}
\label{sec:ana}
\subsection{Two-point functions on masked fields}
Consider two isotropic fields $A^i(\bvec{\Omega})$ and $A^j(\bvec{\Omega})$ defined on the spherical domain $\bvec{\Omega}$.
The direction-independent real-space two-point function, the correlation function\footnote{Outside of cosmology, the correlation function is called covariance function.}, is given by the expectation 
\begin{equation}
\label{eq:def_corr}
    \xi^{ij} (\theta) = \bigl\langle A^i (\bvec{\theta}') \, A^j (\bvec{\theta'} - \bvec{\theta})\bigr\rangle
\end{equation}
over all points with an angular separation $\vert\bvec{\theta}\vert = \theta$ and with respect to realizations of the fields.
It can be understood as an auto- or cross-correlation, depending on whether the same or two different fields are used for the expectation in Eq.~\eqref{eq:def_corr}. 
In practice, the correlation function is estimated on a single field realization using a (weighted) arithmetic mean. A simplified estimator, excluding weights, is given by \citep[e.g.][]{schneider2002a}
\begin{equation}
\label{eq:simp_xi_est}
    \hat{\xi}^{ij} (\theta) = \frac{\textstyle\sum_{kl} A_k^i A_l^j \Delta_{kl} (\theta)}{\textstyle\sum_{kl}\Delta_{kl} (\theta)},
\end{equation}
where points on the fields are labeled by $k$, $l$ and $\Delta_{kl} (\theta)$ selects points separated by $\theta$ or within an angular-separation bin $\Delta \theta$. 
We will denote correlation-function estimators as $\hat{\xi}^{ij} (\theta)$ and introduce another estimator tailored to our use case in Sect.~\ref{sec:corr_func_est}.
It is worth pointing out here that estimators generally estimate a summary statistic based on a sample of finite size of the underlying random variables. In a probabilistic modeling framework, the probability density function, or likelihood, of the estimator in use is crucial, which is why we wish to derive the likelihood for a correlation function estimated on a masked sphere analytically.
The fields in question will be assumed to be Gaussian random fields for the remainder of this work.
We discuss the validity of this assumption in the light of the weak-lensing correlation function in App.~\ref{app:non_linear_scales}.
Nevertheless, the sampling distribution of the correlation-function estimator will in general not be Gaussian.
This can already be seen by looking at Eq.~\eqref{eq:simp_xi_est}, where the Gaussian random variables $A_k$ enter quadratically, which is a non-linear operation and therefore does not retain Gaussianity. 

Only when a sufficient number of independent pairs is considered, i.e. the survey area is large or enough small scale modes are added, will the central limit theorem apply and the distribution of correlation functions will eventually become Gaussian. Generally, no closed form for the full-sky correlation-function likelihood exists \citep{sellentin2018}.

The harmonic space two-point function, the power spectrum ${C}_{\ell}$, can be defined as:
\begin{equation}
\label{eq:def_ps}
    \bigl\langle a^i_{\ell m} \, {a^{\ast \ j}_{\ell' m'}}\bigr\rangle = \delta_{\ell \ell'} \delta_{m m'} C^{ij}_{\ell},
\end{equation}
where the asterisk denotes complex conjugation and the $a_{\ell m}$ are spherical-harmonic-transform coefficients of the respective homogeneous and isotropic Gaussian random fields $A$. The expectation is taken with respect to realizations of the field. From a single field realization, the power spectrum can be estimated as 
\begin{equation}
\label{eq:cell_est}
    \hat{C}^{ij}_{\ell} = \frac{1}{2\ell + 1} \sum_{m} a^i_{\ell m} \, a^{\ast \ j}_{\ell m}.
\end{equation}
Any linear transformation, including the spherical-harmonic transform, retains the Gaussianity of the fields $A$. 
Therefore, the power spectrum estimator is a quadratic form in the Gaussian random variables $a_{\ell m}$. 
If the field is homogeneous and isotropic, the modes $\ell$ are independent and the $a_{\ell m}$ are drawn from a normal distribution with mean zero and variance $\sigma^2 = C_{\ell}$ according to $\ell$ and independently for each $\ell$ and $m$. 
This independence of the individual summands entering the average that represents the estimator sets the power spectrum estimator apart from real-space correlation-function estimators.
In fact, it allows writing down an analytic form of the likelihood and we will therefore use it as a starting point for the correlation-function likelihood.
Explicitly, the likelihood for each mode of the full-sky power spectrum is given by a Gamma distribution or more generally a Wishart distribution, particularly if $i \neq j$ \citep[e.g.][]{hamimeche2008}. 

However, unlike the physical cosmological fields, the observed fields are typically far from homogeneous and isotropic due to the need for a mask or weighting function $W(\bvec{\Omega})$.

The spherical-harmonic-transform coefficients obtained from such a masked field as well as the corresponding estimated power spectrum are therefore described with the prefix ''pseudo'' and should not be confused with the spherical-harmonic-transform coefficients of the underlying cosmological field. 
We will continue with details about this transformation in the next section and will show how a connection to the unmasked equivalents can be made. 

We denote the pseudo-nature with a tilde in equations and write the pseudo-$C_{\ell}$ estimator as the $m$-average of the pseudo-$a_{\ell m}$ 
\begin{equation}
\label{eq:pcell_est}
    \hat{\tilde{C}}^{ij}_{\ell} = \frac{1}{2\ell + 1} \sum_{m} \tilde{a}^i_{\ell m} \, \tilde{a}^{\ast \ j}_{\ell m}.
\end{equation}
Consequently, this is an estimator for the power spectrum of the observed field combining information of the underlying cosmological fields and the mask. 
The key point here is that the inhomogeneity introduced by the mask couples the modes of the pseudo-$a_{\ell m}$ such that they are not independent anymore. 
Therefore, the pseudo-$C_{\ell}$ do not follow the analytic Wishart distribution. 
An exact likelihood for the pseudo-$C_{\ell}$ with arbitrary masks can, albeit not in a closed form, still be obtained \citep{upham2020}.
Combining the known distribution and covariance of the full-sky $a_{\ell m}$ and the mask information, the pseudo-$C_{\ell}$ are suitable to construct a correlation-function estimator, from which we will then continue to derive the likelihood. 
\subsection{Pseudo-\texorpdfstring{$a_{\ell m}$}{alm} for spin-2 fields}
\label{sec:palm}
So far we have not specified the spin of the fields under consideration.
Our example, cosmic shear, can be expressed as a traceless and symmetric rank-2 tensor giving rise to a spin-$2$ field. These can generally be decomposed into $E$- and $B$-modes \citep[see for example][]{bartelmann2010a} and can then be written in terms of the spin-$\pm 2$ spherical harmonics $\tensor[_{\mp 2}]{Y}{_{\ell m}}$ and the $E$- and $B$-mode $a_{\ell m}$ as
\begin{equation}
\label{eq:spin2_exp}
    (\gamma_1 \pm i \gamma_2) (\bvec{\Omega}) = \sum_{\ell, m} (a_{\ell m}^E \mp a_{\ell m}^B) \tensor[_{\mp 2}]{Y}{_{\ell m}}(\bvec{\Omega}),
\end{equation}
where $\gamma_{1/2}$ are the shear fields containing information about galaxy ellipticities and orientations. 
All operations involved in obtaining the $E$- and $B$-mode $a_{\ell m}$ are linear, such that the Gaussianity of the fields implies a Gaussian distribution of the $a_{\ell m}^X$. We will be using capital letters $X$, $Y$ to label different modes of a field and lower case letters $i$, $j$ for different fields (e.g. different tomographic bins). 
This distinction is not strictly necessary because they are all just different fields but it makes the transfer to practical applications a bit easier. 

The pseudo-$a_{\ell m}$ for a spin-$2$ field can be obtained analogously to the full-sky $a_{\ell m}$ from $\tilde{A}(\bvec{\Omega}) = W(\bvec{\Omega}) A(\bvec{\Omega})$ instead of $A(\bvec{\Omega})$. 
A complete derivation can for example be found in \citet{lewis2001a}.
As the multiplication in real space corresponds to a convolution in harmonic space, this effectively introduces a mixing between $E$- and $B$-modes but also between different multipoles $\ell$.
One can show that the pseudo-$a_{\ell m}$ can be written in terms of their full-sky counterparts:
\begin{align}
\label{eq:palm}
    \tilde{a}_{\ell m}^E &= \sum_{\ell', m'} \left( W_{\ell \ell' m m'}^+ a_{\ell'm'}^E + W_{\ell \ell' m m'}^- a_{\ell'm'}^B \right) \\
    \tilde{a}_{\ell m}^B &= \sum_{\ell', m'} \left(W_{\ell \ell' m m'}^+ a_{\ell'm'}^B - W_{\ell \ell' m m'}^- a_{\ell'm'}^E\right), \nonumber
\end{align}
where the mixing matrices $W_{\ell \ell' m m'}^{\pm}$ arise from the spherical-harmonic transform of the weighting map or mask $W(\bvec{\Omega})$.
We refer to App.~\ref{app:mixmat} for a definition of these mixing matrices and will use the analytic form presented in \citet{Hamimeche_2009} for practical implementations.
Equation~\eqref{eq:palm} is a linear combination of the Gaussian random variables $a_{\ell m}^X$ such that the pseudo-$a_{\ell m}^X$ will also be Gaussian distributed but different modes $\ell$ will not be independent anymore.
The spin-$0$ case is slightly simpler but we will not consider it for the remainder of this work. \citet{brown2005} for example also list the spin-$0$ pseudo-$a_{\ell m}$.
\subsection{The correlation function estimator}
\label{sec:corr_func_est}
We have shown that the power spectrum estimator as well as the pseudo power spectrum estimator, Eqs.~\eqref{eq:cell_est} and~\eqref{eq:pcell_est}, are quadratic forms in Gaussian random variables. 
For the pseudo power spectrum the covariance structure can be understood in terms of the full-sky power spectrum.
Conveniently, the correlation function can be expressed using the pseudo-$C_{\ell}$, which we will discuss in this section.

The spin-$2$ pseudo-$C_{\ell}$ estimator can, analogously to the spin-$0$ case, Eq.~\eqref{eq:pcell_est}, compactly be written as
\begin{equation}
    \label{eq:cell}
    \hat{\tilde{C}}_{\ell,(ij)}^{XY} = \tilde{\bvec{a}}^T_{\ell} \bvec{M}^{XY}_{\ell,(ij)} \tilde{\bvec{a}}_{\ell},
\end{equation}
where $\tilde{\bvec{a}}_{\ell}$ is a vector containing all $\tilde{{a}}_{\ell m}$ for a given $\ell$ of all fields ($E$- and $B$-modes, different tomographic bins) under consideration.
The square matrix $\bvec{M}^{XY}_{\ell,(ij)}$ is of order $N_{\mathrm{field}}\ (2\ell + 1)$, where $N_{\mathrm{field}}$ is the number of different fields to correlate. 
This matrix selects the correct $\tilde{a}_{\ell m}$ from the vector $\tilde{\bvec{a}}_{\ell}$ to form the estimator $\hat{\tilde{C}}_{\ell,(ij)}^{XY}$. 
A possible setup of this matrix and way to organize the vector of pseudo-$a_{\ell m}$ has been shown in detail in \citet{upham2020}\footnote{We will adopt a slightly different ordering for the pseudo-$a_{\ell m}$ vector but this is unimportant as long as the same order is kept in the covariance matrix.}.

Having introduced the $E$- and $B$-mode pseudo power spectra as the spin-$2$ harmonic-space two-point functions, we now switch to the real-space correlation function. 
On a spin-$2$ field, two correlation functions, $\xi^+$ and $\xi^-$, can be defined from the shear fields $\gamma_1$ and $\gamma_2$, which we introduced in Sect.~\ref{sec:palm}. 
The fields, $\gamma (\bvec{\Omega}) = (\gamma_1 + i \gamma_2) (\bvec{\Omega})$ and its complex conjugate $\gamma^{\ast}$, are, for each pair of points $\bvec{\theta}$, $\bvec{\theta'}$, rotated onto local bases such that the connecting line becomes the $x$-direction. 
The resulting rotated fields are denoted by $\bar{\gamma}$, $\bar{\gamma}^{\ast}$ and define the correlation function
\begin{equation}
    \xi^{+}_{ij} (\theta) = \bigl\langle \bar{\gamma_i}^{\ast} (\bvec{\theta}') \, \bar{\gamma_j} (\bvec{\theta'} - \bvec{\theta})\bigr\rangle
\end{equation}
for the fields $i$ and $j$.
$\xi^{-}$ and correlation functions for scalar fields can be defined similarly but will be left out here for conciseness.
We limit the demonstration of our results to $\xi^{+}$ because its magnitude is significantly larger for weak lensing and the spin-$2$ case is slightly more complex. 
Apart from that, $\xi^{+}$ is also more sensitive to the large scales we are interested in, as we demonstrate in App.~\ref{app:kernels}.

Considering Eq.~\eqref{eq:spin2_exp}, i.e. the connection between real and spherical-harmonic space through the spin-$2$ spherical-harmonic transform, the corresponding two-point functions can directly be converted between real and harmonic space. 
Based on this, \citet{chon2004} derive a correlation-function estimator in terms of the pseudo-$C_{\ell}$:
They start from a real-space estimator of the continuous masked fields $W(\bvec{\Omega})\bar{\gamma}(\bvec{\Omega})$, make use of the spherical-harmonic transform of the fields and mask function and finally employ the estimator for the pseudo power spectrum, Eq.~\eqref{eq:pcell_est}.

We will be using the resulting correlation-function estimator for the remainder of this work and it reads
\begin{equation}
\label{eq:xip}
    \hat{\xi}^{+}_{ij} (\theta) = 2 \pi B(\theta) \sum_{\ell} \ (2 \ell + 1) d_{2 2}^{\ell} (\theta) (\tilde{C}_{\ell,(ij)}^{EE} + \tilde{C}_{\ell,(ij)}^{BB}),
\end{equation}
where  $d_{2 2}^{\ell} (\theta)$ is a Wigner $d$-function coming from spatial integrals over products of spin-weighted spherical harmonics \citep[for a definition of the Wigner (small) $d$-matrices, see][]{goldberg1967a,tessore2019a}.
We also specified $E$- and $B$-mode combinations for the pseudo-$C_{\ell}$. 
Again, lower case Roman letters refer to different fields that could be arising from different redshift distributions. $B(\theta)$ is a normalization factor:
\begin{equation}
\label{eq:norm}
    B(\theta) = \left(2 \pi \sum_{\ell} \left(2 \ell + 1 \right) P_{\ell} \left(\cos \theta \right) w_{\ell}\right)^{-1},
\end{equation}
where $P_{\ell}$ are Legendre polynomials and $w_{\ell}$ denotes the (spin-$0$) mask power spectrum. The reciprocal of Eq.~\eqref{eq:norm} is essentially the mask correlation function.

It is important to note that we will always be taking the sum over multipole moments in this normalization factor to the bandlimit of a given field to ensure convergence.
This will be completely consistent once we also apply the high-$\ell$ extension to our likelihood (see Sect.~\ref{sec:full}).

Correlation functions are rarely measured for a single angular separation. Instead, galaxy pairs are collected from angular-separation bins.
To include this in our prediction, our estimator is angular-bin integrated, weighted with the angular separation to account for an increasing number of galaxy pairs with increasing angular separation.
This weighting is an approximation that neglects boundary effects introduced by the mask and assumes a uniform galaxy density but could be replaced by a more sophisticated weighting scheme \citep[e.g.][]{asgari2021a}. 
Including the angular-bin integration, the correlation-function estimator becomes
\begin{equation}
\label{eq:xip_cell}
    \hat{\xi}^{+}_{ij} \left(\bar{\theta}\right) =  \sum_{\ell} (2 \ell + 1) K_{\ell} \left(\bar{\theta}\right) \left(\tilde{C}_{\ell,(ij)}^{EE} + \tilde{C}_{\ell,(ij)}^{BB}\right),
\end{equation}
where the angular dependence is absorbed into $K_{\ell} \left(\bar{\theta}\right)$ as
\begin{equation}
\label{eq:k_ell}
    K_{\ell} \left(\bar{\theta}\right) = 2 \pi \  \ \frac{2}{\theta_{\mathrm{max}}^2 - \theta_{\mathrm{min}}^2} \int_{\theta_{\mathrm{min}}}^{\theta_{\mathrm{max}}} \dd \theta \ \theta B(\theta) d_{2 2}^{\ell} (\theta).
\end{equation}
We will always refer to an angular-separation bin as the closed interval $\bar{\theta} = [\theta_{\mathrm{min}},\theta_{\mathrm{max}}]$.
Finally, plugging Eq.~\eqref{eq:pcell_est} into Eq.~\eqref{eq:xip_cell}, we can write
\begin{equation}
    \label{eq:xip_alm}
        \hat{\xi}^{+}_{ij} \left(\bar{\theta}\right) =  \sum_{\ell} K_{\ell} \left(\bar{\theta}\right)  \sum_{m=-\ell}^{\ell} \left(\tilde{a}^{E,i}_{\ell m} {\tilde{a}^{\ast \ E,j}_{\ell m}} +  \tilde{a}^{B,i}_{\ell m} {\tilde{a}^{\ast \ B,j}_{\ell m}}\right).
\end{equation}
All prefactors including the angular-bin integrals can now, as in Eq.~\eqref{eq:cell}, be absorbed into a matrix, such that
\begin{equation}
    \label{eq:xip_M_bin}
    \hat{\xi}^{+}_{ij} (\bar{\theta}) = \tilde{\bvec{a}}^T \bvec{M}_{ij}^{\xi^{+}}(\bar{\theta}) \tilde{\bvec{a}}.
\end{equation}
We will later write the complex pseudo-$a_{\ell m}$ in terms of their real and imaginary parts as these are distributed independently.

Now, the $ \tilde{\bvec{a}}$ vectors are just the $\tilde{\bvec{a}}_{\ell}$ vectors stacked, resulting in a length of $N_{\mathrm{field}} \sum_{\ell = \ell_{\mathrm{min}}}^{\ell_{\mathrm{max}}} (2 \ell + 1) = N_{\mathrm{field}} (\ell_{\mathrm{max}}^2 - \ell_{\mathrm{min}}^2 + 2 \ell_{\mathrm{max}} + 1)$ and the matrix $\bvec{M}_{ij}^{\xi^{+}} (\theta)$ is a diagonal matrix taking care of the double sum.
This is again a quadratic form in the pseudo-$a_{\ell m}$. 
\subsection{Characteristic function}
\label{sec:cf_der}
As shown in the previous section, any two-point function, and particularly the correlation function we are mainly interested in, can be written as a quadratic form. 
Quadratic forms of Gaussian random variables have a known characteristic function. 
Characteristic functions are the Fourier counterpart to probability density functions (PDF) and we therefore introduce the Fourier pair $\{\bvec{\hat{\xi}},\bvec{t} \}$, for an $n$-dimensional joint likelihood for $n$ correlation-function measurements $\bvec{\hat{\xi}}$. 

In practice, the characteristic function will be defined on a discrete grid $\bvec{t}$ with $(N_{\mathrm{grid}})^n$ grid points, where $N_{\mathrm{grid}}$ is the number of gridpoints along one dimension that needs to be specified to be large enough to resolve the behavior of the characteristic function. 
Meanwhile, the range of $\bvec{t}$ needs to be adjusted to cover the support of the corresponding PDF of the correlation-function measurement. This is accomplished by setting the spacing of the gridpoints to the inverse of the largest expected correlation function value for each dimension.

The characteristic function for quadratic forms of Gaussian random variables defined on $\bvec{t}$ can be written in several ways.
We will be using one way of writing it for numerical computational purposes and one for analytical considerations. 
For analytical calculations, we use the form presented in \citet{good1963}:
\begin{equation}
\label{eq:char_ana}
    \varphi (\bvec{t}) = \vert \mathbb{1} - 2i \sum_k t_k \bvec{M}_k \bvec{\Sigma} \vert^{-1/2},
\end{equation}
where $\bvec{\Sigma}$ is the covariance matrix of the vector of Gaussian random variables, the pseudo-$a_{\ell m}$ in our case, and $\bvec{M}_k$ are square matrices combining the pseudo-$a_{\ell m}$ to yield the estimators $\bvec{\hat{\xi}}$ in question as shown in Eq.~\eqref{eq:xip_M_bin}.
Gridpoints in $\bvec{t}$ are described by tuples with elements $t_k$.
\citet{upham2020} showed that this characteristic function can equivalently be written as
\begin{equation}
\label{eq:char_num}
    \varphi (\bvec{t}) = \prod_j (1-2i\lambda_j)^{-1/2},
\end{equation}
where the $\lambda_j$ are eigenvalues of the product of characteristic function variables, combination matrices, and the covariance matrix as used in Eq.~\eqref{eq:char_ana}:
\begin{equation}
\label{eq:eigvals}
    \lambda_j \in \lambda \left(\sum_k t_k \bvec{M}_k \bf{\Sigma}\right).
\end{equation}

For the one-dimensional case, i.e. for the likelihood of a measurement of a single correlation function value, the gridpoints $t$ of the one-dimensional grid can be pulled out of the sum. 
The characteristic function for each gridpoint in the one-dimensional case is given by
\begin{equation}
    \varphi (t) = \prod_j (1-2i t \lambda_j)^{-1/2},
\end{equation}
where the $\lambda_j$ only need to be computed once as 
\begin{equation}
\label{eq:eigvals_1d}
    \lambda_j \in \lambda \left(\bvec{M} \bf{\Sigma}\right).
\end{equation}

The latter way of writing the characteristic function, Eq.~\eqref{eq:char_num}, is easier to treat numerically, which is why we followed \citet{upham2020} in implementing it in our framework.

The likelihood follows from the characteristic function as $p(\bvec{\hat{\xi}} \vert \bvec{\vartheta}) \propto \int \dd \bvec{t} \ \varphi(\bvec{t}, \bvec{\vartheta}) \ \mathrm{e}^{-i \bvec{t}\bvec{\hat{\xi}}}$, where we explicitly added the dependence of the characteristic function on the cosmological parameters $\bvec{\vartheta}$, which come in through the covariance matrix $\bvec{\Sigma}(\bvec{\vartheta})$. 

\subsubsection{Covariance matrix of the pseudo-\texorpdfstring{$a_{\ell m}$}{alm}}
Having shown in Sect.~\ref{sec:palm} that the pseudo-${a}_{\ell m}$ are Gaussian distributed if the fields and therefore the full-sky ${a}_{\ell m}$ are, we will now determine their covariance matrix.
This is the covariance matrix that appears in Eq.~\eqref{eq:char_ana}, the characteristic function of the correlation-function likelihood.
More precisely, it is the covariance matrix for a set of pseudo-${a}_{\ell m}$, i.e. $E$- and $B$-modes, needed for the case of the $\xi^+$-correlation-function estimator, Eq.~\eqref{eq:xip_alm}.
For the likelihood of a single auto-correlation function, we need the covariance matrix of pseudo-${a}_{\ell m}$ belonging to one set of full-sky power spectra $C^{XY}_{\ell, ii}$.
Considering cross-correlations across redshift bins, the auto- and cross-full-sky power spectra and pseudo-${a}_{\ell m}$ covariances for both redshift bins are required -- we will come back to this later.
Each pseudo-${a}_{\ell m}$ can be written as linear combination of ${a}_{\ell m}$, Eq.~\eqref{eq:palm}. 
This can be generalized to 
\begin{equation}
   (\tilde{a}_{\ell m})_{\alpha} = \sum_{\ell', m',\sigma \in \Phi({\alpha})} (W_{\ell \ell' m m'})_{\sigma} (a_{\ell'm'})_{\sigma},
\end{equation}
where the left-hand side can now either be a real or imaginary part and belong to any field. The selection of real or imaginary part and $E$- or $B$-mode is encapsulated by $\alpha$.
We introduced a mapping $\Phi \colon \alpha \mapsto \Phi(\alpha)$ for the selection of the right $W_{\ell \ell' m m'}$-$a_{\ell m}$-pairs in the linear combination, i.e., each index $\alpha$ is mapped to four sets of real or imaginary parts of $W_{\ell \ell' m m'}^{\pm}$, $a_{\ell m}^{E/B}$ and four corresponding signs $\pm 1$.
For example, $\alpha = 0$ selects the real part of the $E$-mode pseudo-$a_{\ell m}$, such that
\begin{equation}
\begin{split}
       (\tilde{a}_{\ell m})_{\alpha=0} &= \operatorname{Re}(\tilde{a}^E) \\ &= \operatorname{Re}({W}^+)\operatorname{Re}({a}^E) + \operatorname{Re}({W}^-)\operatorname{Re}({a}^B) \\
       &\phantom{=} \quad - \operatorname{Im}({W}^+)\operatorname{Im}({a}^E) - \operatorname{Im}({W}^-)\operatorname{Im}({a}^B),
\end{split}
\end{equation}
where we dropped all indices and sums for clarity but summation over $\ell'$ and $m'$ is still implied. 

Projecting a complex variable onto its real or imaginary part again is a linear operation, therefore these parts are still Gaussian distributed.
This means that each component of the pseudo-${a}_{\ell m}$ amounts to a sum over products of some part of a field-specific but constant mixing matrix $W_{\ell \ell' m m'}^{\pm}$ and some part of a spherical-harmonic coefficient of the underlying Gaussian random field.
The covariances of the full-sky ${a}_{\ell m}$ are known and given by the angular power spectrum as defined in Eq.~\eqref{eq:def_ps}.

The covariance of linear combinations of random variables can be obtained from the known covariances of the latter, such that
\begin{equation}
\begin{split}
\operatorname{cov}((\tilde{a}_{\ell m})_{\alpha},(\tilde{a}_{\ell' m'})_{\beta}) &= \\ 
\sum_{\ell'', m''} \sum_{{\ell''', m'''}} \sum_{{\substack{\sigma \in \Phi({\alpha}), \\ \kappa \in \Phi({\beta})}}} &(W_{\ell \ell'' m m''})_{\sigma} (W_{\ell' \ell''' m' m'''})_{\kappa} \\
&  \times \operatorname{cov}((a_{\ell''m''})_{\sigma},(a_{\ell'''m'''})_{\kappa}).
\end{split}
\end{equation}
Conveniently, the covariance for ${a}_{\ell m}$ with different $\ell$ vanishes, such that a part of the second sum can be eliminated:
\begin{equation}
\label{eq:cov_all}
\begin{split}
    \operatorname{cov}((\tilde{a}_{\ell m})_{\alpha},(\tilde{a}_{\ell' m'})_{\beta}) &= \\
    \sum_{\ell'', m''} \sum_{m'''} \sum_{\substack{\sigma \in \Phi({\alpha}), \\ \kappa \in \Phi({\beta})}} &(W_{\ell \ell'' m m''})_{\sigma} (W_{\ell' \ell'' m' m'''})_{\kappa} \\
    & \times \operatorname{cov}((a_{\ell''m''})_{\sigma},(a_{\ell''m'''})_{\kappa}).
\end{split}
\end{equation} 
We can see here explicitly that the covariance of pseudo-${a}_{\ell m}$ of different $\ell$ does not vanish anymore, unlike the covariance of full-sky ${a}_{\ell m}$.
That is, the pseudo-${a}_{\ell m}$ are still Gaussian distributed with zero mean but they are not independent.

The $W_{\ell \ell'' m m''}^{\pm}$ can be precomputed and are saved in a five-dimensional complex array - four dimensions for all possible combinations of $(\ell \ell'' m m'')$ and the fifth for the choice of the linear combination, $+$ or $-$.
The covariances have to be assigned in the following way: 
\begin{multline}
    \operatorname{cov}((a_{\ell''m''})_{\sigma},(a_{\ell''m'''})_{\kappa})) = \\
    \medmath{\begin{cases}
        \frac{1}{2} C_{\ell''}^{\sigma \kappa}, & \text{if both $a_{\ell''m''}$ are real or imaginary and $m'' = m'''$} \\
        (-1)^m  \frac{1}{2} C_{\ell''}^{\sigma \kappa}, & \text{if both $a_{\ell''m''}$ are real and $m'' = -m'''$} \\
        (-1)^{m+1}  \frac{1}{2} C_{\ell''}^{\sigma \kappa}, & \text{if both $a_{\ell''m''}$ are imaginary and $m'' = -m'''$} \\
        C_{\ell''}^{\sigma \kappa}, & \text{if both $a_{\ell''m''}$ are real and $m'' = 0$} \\
        0, & \text{otherwise}
    \end{cases}},
\end{multline}
where the factor $1/2$ arises due to the fact that we consider covariances of real and imaginary parts of the ${a}_{\ell m}$ separately.\footnote{The ${a}_{\ell 0}$ are real due to the property ${a}_{\ell -m} = (-1)^m {a}_{\ell m}^{\ast}$.}

Note that Greek letters label fields but also real or imaginary parts, so they determine the covariance by selecting the appropriate $C_{\ell}$ but also have an influence on the sign. 
Continuing the $\alpha=0$ example from above, this $\Phi(\alpha = 0)$ would select a subset such that $\sigma$ iterates over $\operatorname{Re}({a}^E), \operatorname{Re}({a}^B), \operatorname{Im}({a}^E), \operatorname{Im}({a}^B)$ and corresponding $W_{\ell \ell'' m m''}^{\pm}$ and if we also choose $\beta =0$, the same elements would be selected for $\kappa$ via $\Phi(\beta=0)$.
For the covariances, we would then sum over the corresponding combinations.
Written out versions of this index notation for the pseudo-${a}_{\ell m}$ covariance matrix can be found in \citet{upham2020}.

Computing the characteristic function for the correlation-function estimator, Eq.~\eqref{eq:xip_alm}, requires the complete covariance matrix for all pseudo-${a}_{\ell m}$, i.e. $E$- and $B$-modes as well as real and imaginary parts, as they all appear in this estimator. In practice, the sums in Eq.~\eqref{eq:cov_all} are taken to slightly higher multipole moments than needed for the covariance matrix to allow for a margin for convergence, as the mixing matrices involve the oscillatory Wigner $d$-functions.
\subsubsection{Combination matrices}
The characteristic function can be calculated from the covariance matrix explained in the previous section and a combination matrix as introduced in Eq.~\eqref{eq:xip_M_bin} for the correlation function we are interested in.
For the one-dimensional auto-correlation and one of the pseudo-${a}_{\ell m}$ modes, this matrix can be written out as
\begin{equation}
    \label{eq:m_onemode}
\bvec{M} \left(\bar{\theta}\right) =
    \begin{pmatrix}
        K_0 \left(\bar{\theta}\right)& 0 & \dots & & \dots & 0\\
        0 & \ddots &  & & & \vdots \\
        \vdots &  & K_{\ell_{\mathrm{max}}} \left(\bar{\theta}\right) &  & & \\
         &  & & K_0 \left(\bar{\theta}\right) & &\vdots \\
        \vdots &  & & & \ddots & 0\\
        0 & \dots & & \dots & 0 & K_{\ell_{\mathrm{max}}} \left(\bar{\theta}\right)
    \end{pmatrix},
\end{equation}
where each $K_{\ell} \left(\bar{\theta}\right)$, Eq.~\eqref{eq:k_ell}, is repeated for each $m$.\footnote{In practice, $K_{\ell} \left(\bar{\theta}\right)$ is repeated $(\ell_{\mathrm{max}} + 1)$ times, as only positive $m$ are required when using real and imaginary parts. The padding up to $\ell_{\mathrm{max}}$ instead of just repeating up to the current $\ell$ is implemented for practical computational reasons. This structure is also applied to the pseudo-${a}_{\ell m}$ covariance matrix.}
This sequence is written out twice, once for each real and imaginary part, resulting in a $ 2\ (\ell_{\mathrm{max}} + 1)^2$ dimensional matrix.
The angular dependence is only in these combination matrices such that the same covariance matrix of the pseudo-${a}_{\ell m}$ can be used for all angular-separation bins.
For the full correlation function, $\bvec{M} \left(\bar{\theta}\right)$ is repeated twice along the diagonal, once for the $E$- and once for the $B$-mode:
\begin{equation}
    \label{eq:m_auto}
\bvec{M}^{\xi^{+}}_{ii} \left(\bar{\theta}\right) =
\begin{pmatrix}
    \bvec{M} (\bar{\theta}) & 0 \\
     0 &  \bvec{M} (\bar{\theta}).
\end{pmatrix}.
\end{equation}
An example for a cross-correlation combination matrix can be found in App.~\ref{app:comb_mat}.

To build a joint likelihood of multiple correlation function measurements, a combination matrix has to be set up for each dimension in such a way that the same order of pseudo-${a}_{\ell m}$ and therefore covariance matrix can be used for each summand according to Eq.~\eqref{eq:eigvals}.

\subsection{Obtaining the likelihood}
\label{sec:likelihood_charac}
Having calculated the pseudo-${a}_{\ell m}$ covariance matrix and constructed the combination matrix or matrices, obtaining the characteristic function requires the retrieval of the eigenvalues of the product of the two. For the one-dimensional case, this only needs to be done once and the resulting eigenvalues can straightforwardly be multiplied with the characteristic function grid parameters $t$.
There are two computational limitations in that. 
First, going to higher multipole moments the sidelengths $N$ of the covariance and combination matrices scale with the maximum multipole moment squared  times the number of fields to correlate, such that $N = \mathcal{O}(N_{\mathrm{field}}\ell_{\mathrm{max}}^2)$.
As computing the eigenvalues is always an $\mathcal{O}(N^3)$ operation, this becomes computationally intractable for large $N_{\mathrm{field}}$ and $\ell_{\mathrm{max}}$. 
We will get back to this issue in Sect.~\ref{sec:full}. 
Second, for $n$-dimensional likelihoods, these eigenvalues need to be retrieved $(N_{\mathrm{grid}})^n$ times as each gridpoint becomes a tuple $t_k$ that cannot be taken out of the sum in Eq.~\eqref{eq:eigvals}.

Once the characteristic function is calculated on the whole grid however, retrieving the likelihood is a straightforward Fourier transform that can be obtained quickly and on $n$-dimensional grids using standard libraries.

Useful properties of the PDF, such as its moments, can be obtained analytically without having to calculate the characteristic function on the full grid explicitly. 
We will showcase this for the first two moments in the following. 
\subsubsection{Derivation of the mean}
\label{sec:ana_mean}
The moments of a random variable $\xi$ can be derived from its characteristic function as
\begin{equation}
    \varphi_{\xi}^{(k)}(0) = i^k \operatorname{E}[\xi^k],
\end{equation}
where $k$ numbers the moments (i.e. $k=1$ for $\mu_1 = \mu$ the mean, $k=2$ for $\mu_2 = \mu^2 + \sigma^2$ containing the variance, ect.) and the superscript $(k)$ denotes the $k$-th derivative with respect to the characteristic function parameter, evaluated at zero.
Using the closed form representation of our characteristic function, Eq.~\eqref{eq:char_ana}, and making use of Jacobi's formula, the mean of a quadratic form of Gaussian random variables with combination matrix $\bvec{M}$ and covariance matrix $\bvec{\Sigma}$ is given by
\begin{equation}
    \label{eq:lowmean}
    \operatorname{E}[\xi] =  \mathrm{tr} \left(\bvec{M} \bvec{\Sigma}\right).
\end{equation}
$\bvec{M}$ is a diagonal matrix in the auto-correlation case, see Eq.~\eqref{eq:m_auto}, such that the product becomes a row-wise multiplication of $\bvec{\Sigma}$ with the entries of $\bvec{M}$, which are the $K_{\ell} \left(\bar{\theta}\right)$, Eq.~\eqref{eq:k_ell}. 
The trace selects the auto-correlations of all pseudo-${a}_{\ell m}$, the corresponding variance, so we can write
\begin{equation}
    \mathrm{tr} \left(\bvec{M} \bvec{\Sigma}\right) = \sum_{\ell,m,X} K_{\ell} \left(\bar{\theta}\right) \left(\operatorname{var} (\operatorname{Re}(\tilde{a}_{\ell m}^X)) + \operatorname{var} (\operatorname{Im}(\tilde{a}_{\ell m}^X))\right),
\end{equation}
where the sum is taken over all $\ell$, $m$ and modes $X$.
This is due to the way the covariance matrix $\bvec{\Sigma}$ is set up and how the trace acts on it.
The variances of the pseudo-${a}_{\ell m}$ are given by their squared $m$-averages, for example 
\begin{equation}
    \operatorname{var} (\operatorname{Re}(\tilde{a}_{\ell m}^X)) = \frac{1}{2\ell + 1} \sum_{m=-\ell}^{\ell} \operatorname{Re}(\tilde{a}_{\ell m}^X)^2
 \end{equation}
 for the real parts. 
 The variance for the imaginary parts is obtained in the same way.\footnote{The imaginary part of the $m=0$ component is zero and the pseudo-$C_{\ell}$ estimator can also be written down using only the positive $m$, which we practically do in the implementation, but we express both variances the same way without explicitly taking this into account here for consistency with our previous definitions.}
The sum of these two components (variances of real and imaginary parts) amounts to the pseudo-$C_{\ell}$ estimator, Eq.~\eqref{eq:pcell_est}, resulting in $\tilde{C}_{\ell}^{XX}$. 
Additionally performing the sum over $m$ giving a factor of $(2\ell + 1)$ because the $\tilde{C}_{\ell}$ are independent of $m$, we arrive at the mean
\begin{equation}
    \mathrm{tr} \left(\bvec{M} \bvec{\Sigma}\right) = \sum_{\ell,X} K_{\ell} \left(\bar{\theta}\right) (2\ell + 1) \, \tilde{C}_{\ell}^{XX},
\end{equation}
which is exactly the correlation function estimator, Eq.~\eqref{eq:xip_cell}. 
The mean of the PDF will therefore correspond to the estimator itself, as it should. Similar arguments hold for the cross-correlation case, where the matrix multiplication in Eq.~\eqref{eq:lowmean} can be performed block-wise.
\subsubsection{Derivation of the variance}
Similarly to the mean, the second moment of a given PDF can be written in terms of its characteristic function as
\begin{equation}
    \operatorname{E}[\xi^2] = -  \varphi_{\xi}^{(2)}(0).
\end{equation}
For our particular characteristic function, this yields the variance
\begin{equation}
\label{eq:exact_var}
    \sigma^2 = 2 \mathrm{tr} \left(\bvec{M} \bvec{\Sigma}\bvec{M} \bvec{\Sigma}\right).
\end{equation}
A similar expression can also be derived by starting from Isserlis's identity \citep[][]{isserlis1916a} for the covariance of a quadratic estimator, as \citet{dahlen2008a} demonstrate for the power spectrum. 
To see whether the characteristic function has been set up properly we compare the result of Eq.~\eqref{eq:exact_var} to the variance of the exact likelihood by integration of the obtained real-space PDF.

\section{Likelihood of the Full Correlation Function}
\label{sec:full}
As the sidelength $N$ of the covariance matrix scales with the maximum multipole moment used for the calculation of the correlation function as $N = \mathcal{O}(\ell_{\mathrm{max}}^2)$, the computational effort increases accordingly; eigenvalues need to be retrieved from these matrices to obtain the characteristic function. 
The retrieval of the eigenvalues itself is an $\mathcal{O}(N^3)$ operation such that it will scale with the maximum multipole moment as $\mathcal{O}(\ell_{\mathrm{max}}^6)$. 
This makes calculating a likelihood for the correlation function up to the bandlimit of a given mask essentially infeasible. 
However, as can be seen from the correlation function estimator, Eq.~\eqref{eq:xip_alm}, $2 \ell + 1$ sets of pseudo-$a_{\ell m}$ are added for each $\ell$.
Therefore, an increasing number of modes is added when going to higher $\ell$.
As the squared pseudo-$a_{\ell m}$ are random variables, their sum tends to become a Gaussian distribution due to the central limit theorem.\footnote{In many versions of the central limit theorem, independence of the summands is assumed, which is not the case for the pseudo-$a_{\ell m}$. Empirically, it seems to apply nevertheless.}
\cite{hall2022a} have explored the transition to the Gaussian regime for the full-sky power spectrum, even in the presence of mild non-Gaussianities in the field, and found that the likelihood converges to a Gaussian distribution on small scales. Our correlation-function estimator, as expressed in Eq.~\eqref{eq:xip}, is a sum of pseudo-$C_{\ell}$. Consequently, we expect the individual components, and even more so a partial sum over these components, to approach a Gaussian distribution beyond a certain threshold $\ell_{\mathrm{exact}}$.
Based on this reasoning, we assume that the sampling distribution of the high-$\ell$ part of the sum in the correlation-function estimator as expressed in Eq.~\eqref{eq:xip} for large enough $\ell$ can be approximated by a Gaussian distribution. We therefore split the sum at a cutoff $\ell_{\mathrm{exact}}$.
The characteristic functions of each of these two random variables, the (exact) low-multipole-moment part and the approximate high-multipole-moment part, can then be calculated separately.
Using the relationship between characteristic functions and their PDFs, adding the two random variables then becomes a point-wise multiplication in the characteristic-function space. 
Doing so, we implicitly assumed that the two parts of the sum are independent. 
This is not necessarily true, as all pseudo-$C_{\ell}$ are correlated \citep[see][]{upham2020}.
However, we find that the assumption of independence does not affect our empirical results.
We will also find that the assumption of the applicability of the central limit theorem can be justified by showing that the correlation-function likelihood converges to the simulated sampling distributions as the cutoff is shifted to higher $\ell_{\mathrm{exact}}$.
The full characteristic function becomes
\begin{equation}
    \varphi (\mathbf{t}) = \varphi_{\mathrm{exact}} (\mathbf{t})  \varphi_{\mathrm{Gauss}} (\mathbf{t}),
\end{equation}
where $\varphi_{\mathrm{exact}}$ is given by the exact characteristic function as discussed in Sect.~\ref{sec:cf_der} calculated up to a cutoff $\ell_{\mathrm{exact}}$ and the Gaussian part, $\varphi_{\mathrm{Gauss}}$ is characterized by its mean $\mathbf{\mu}$ and covariance matrix $\mathbf{\Sigma}_{\mathrm{Gauss}}$. 

The means were taken to be the corresponding partial sums over pseudo-$C_{\ell}$ from the correlation function estimator, Eq.~\eqref{eq:xip}:
\begin{equation}
\label{eq:gauss_mu}
    \mu = 2 \pi B(\theta) \sum_{\ell = \ell_{\mathrm{min}}}^{\ell_{\mathrm{max}}} \ (2 \ell + 1) d_{2 2}^{\ell} (\theta) (\tilde{C}_{\ell}^{EE} + \tilde{C}_{\ell}^{BB}).
\end{equation}
For this application, we analytically computed the pseudo-$C_{\ell}$ following equation~$9$ in \citet{alonso2019a} instead of summing over pseudo-$a_{\ell m}$. 
In Eq.~\eqref{eq:gauss_mu}, $\ell_{\mathrm{max}}$ refers to the \texttt{HEALPix}\footnote{\url{http://healpix.sf.net}} \citep{gorski2005a,zonca2019a} bandlimit of a map with the same resolution as the involved mask given by $3N_{\mathrm{side}} - 1$ and the lower limit of the sum depends on the maximum multipole moment the exact calculation is taken to such that $ \ell_{\mathrm{min}} =  \ell_{\mathrm{exact}} + 1$.
As the mean of the sum of two random variables is given by the sum of their means, Eq.~\eqref{eq:gauss_mu} and the low-multipole-moment part with mean Eq.~\eqref{eq:lowmean} add up to the unbiased mean given by the estimator, Eq.~\eqref{eq:xip}.

We estimate the Gaussian covariance for the correlation function following~\citet{joachimi2021a}. 
We replace the integral over the angular power spectrum times the Bessel function by a discrete sum over integrated Wigner $d$-functions in order to consistently use the full-sky formalism.
The influence of the mask on the covariance will be approximated by using an effective sky fraction $f_{\mathrm{sky}}$. 
Using the $f_{\mathrm{sky}}$ approximation instead of the actual weights of the map seems to be sufficient for the small scales the high-$\ell$ part of the likelihood is applied to. 
We note in passing that a more sophisticated estimate of the correlation-function covariance for the high-$\ell$ part that for example takes into account the precise mask geometry is possible  \citep{garcia-garcia2019a,nicola2021a} but does not seem to be necessary at this point. 
The Gaussian covariance of the high-$\ell$ parts of the cosmic-shear correlation functions $\xi^+_{ij}(\bar{\theta}_1)$ and $\xi^+_{kl}(\bar{\theta}_2)$, $\bvec{\Sigma}_{\mathrm{Gauss}}$, therefore can be written out as
\begin{equation}
    \label{eq:gcov_xi}
    \begin{split}
    \operatorname{cov} \left(\xi^+_{ij}(\bar{\theta}_1),\xi^+_{kl}(\bar{\theta}_2)   \right) &= \\ 
    \frac{1}{f_{\mathrm{sky}}} \sum_{\ell =\ell_{\mathrm{min}}}^{\ell_{\mathrm{max}}} & \mathcal{K}_{\ell}(\bar{\theta}_1) \mathcal{K}_{\ell}(\bar{\theta}_2) \left(C_{\ell}^{(ik)} C_{\ell}^{(jl)} + C_{\ell}^{(il)} C_{\ell}^{(jk)}\right)
    \end{split}
\end{equation}
where $i$, $j$, $k$ and $l$ refer to tomographic bins and $\bar{\theta}_{1/2}$ label the corresponding angular-separation bins.
The $C_{\ell}$ terms in Eq.~\eqref{eq:gcov_xi} are only the $E$-modes, as, to first order and ignoring the effects of intrinsic alignment (see also Section~\ref{sec:wlapp}), the $B$-modes are zero in weak lensing \citep[][]{schneider1998a,hilbert2009a}.
In the implementation, the $B$-modes are added in the same way.
This is because we assume the $C_{\ell}$ in this equation to already include the corresponding noise power spectrum and there is an (equal) shot-noise contribution to both parts of this spin-$2$ signal.
We will discuss the shot noise in more detail in Sect.~\ref{sec:noise}.  

The $\mathcal{K}_{\ell}$ are essentially the angular-separation bin-integrated Wigner $d$-functions and other prefactors:
\begin{equation}
    \mathcal{K}_{\ell} (\bar{\theta}) = \frac{2\ell + 1}{4 \pi} \frac{2}{\theta_{\mathrm{max}}^2 - \theta_{\mathrm{min}}^2} \int_{\theta_{\mathrm{min}}}^{\theta_{\mathrm{max}}} \dd \theta \ \theta d_{2 2}^{\ell} (\theta).
\end{equation}

Finally, the characteristic function of this multivariate normal distribution is given by $\varphi_{\mathrm{Gauss}} (\mathbf{t}) = e^{i{ \mathbf{t}^{\mathrm{T}} \boldsymbol{\mu}}-\frac {1}{2} \mathbf{t}^{\mathrm{T}}\boldsymbol{\Sigma}_{\mathrm{Gauss}} \mathbf{t}}$.
Next to the use as a high-$\ell$ extension for the exact likelihood, we will also use this definition of the Gaussian covariance $\Sigma_{\mathrm{Gauss}}$, Eq.~\eqref{eq:gcov_xi}, and the mean $\mu$, Eq.~\eqref{eq:gauss_mu}, to compare our exact likelihood to a Gaussian likelihood in Sect.~\ref{sec:gausscomp}, for which we will set $\ell_{\mathrm{min}} = 2$.

\section{Application to Weak Lensing Correlation Functions}
\label{sec:wlapp}
To show the applicability of our theoretical likelihood, we calculate correlation-function likelihoods for a weak-lensing survey setup.
The prediction of the field values for this observable is obtained by projecting the three-dimensional matter density distribution onto the sphere by performing a weighted integral along the line of sight. 
This procedure approximately follows the paths of photons that are emitted by a distribution of source galaxies $n_i(z)$ and trace the geometry of spacetime via the gravitational potential, which is sourced by the matter distribution between us and the source galaxies \citep[for a review, see][]{bartelmann2001a}. 
The resulting projected field of initially randomly distributed galaxy shapes and orientations\footnote{This is a simplifying assumption as galaxies can be intrinsically aligned due to the large scale structure \citep[for an overview, see][]{joachimi2015a}.} can be correlated with itself or with a field arising from a different source-redshift distribution.
Translating the resulting correlation function into harmonic space, the two-dimensional shear power spectrum $C_{\ell}$ can be constructed as a projection of the three-dimensional matter power spectrum.
As a similar matter distribution is seen by photons emitted from galaxies close in angular separation, the observed galaxy shapes and orientations will become correlated due to this weak-lensing effect.

The three-dimensional power spectrum of the matter distribution can be predicted from the cosmological model, while the redshift distributions of the source galaxies need to be estimated. 
We chose the source redshift bins $S5$ and $S3$ of the KiDS-1000 survey with redshift distributions $n_5(z)$ and $n_3(z)$, roughly corresponding to redshift ranges of $z_5 = [0.9,1.2]$ and $z_3 = [0.5,0.7]$.
Details about these redshift distributions can be found in \citet{hildebrandt2021a}.
We computed our theory power spectra for cosmic shear using these redshift distributions and a flat cold-dark-matter model with a cosmological constant ($\Lambda$CDM) from the Core Cosmology Library\footnote{\url{https://github.com/LSSTDESC/CCL}} \citep[CCL;][]{chisari2019a}. 
The model parameters were set to $\Omega_{\mathrm{c}}=0.25$ for the cold dark matter density, $\Omega_{\mathrm{b}} = 0.05$ for the baryon density, $\sigma_8 = 0.81$ for the variance of the matter density fluctuations, $n_{\mathrm{s}} = 0.96$ for the primordial scalar perturbation spectral index, and $h = 0.67$ for the Hubble constant divided by $\SI{100}{\kilo \metre \per \second \per \mega \parsec}$, if not indicated differently (only relevant in Sect.~\ref{sec:post}).

In using the estimator Eq.~\eqref{eq:xip_alm} and these theory power spectra, we kept the Limber-approximation but let go of the flat-sky approximation. 
Particularly on large scales, the flat-sky approximation is insufficient, whereas the Limber-approximation used to project the three-dimensional matter power spectrum onto the two-dimensional spherical power spectrum can be justified for the broad weak-lensing kernels used in the integration \citep{Lemos_2017,gao2023a}.
However, applying our exact likelihood to data from galaxy surveys would require revisiting this assumption, particularly on the large scales we are interested in \citep[][]{leonard2023a,chiarenza2024a,reymond2025a}. 

Depending on how many redshift bins shall be correlated, the pseudo-${a}_{\ell m}$ covariance matrix needs to be calculated for several sets of pseudo-${a}_{\ell m}$ according to Eq.~\eqref{eq:cov_all} using the corresponding theory power spectra for all different redshift-bin combinations.

The combination matrix $\bvec{M}_{ij}^{\xi^{+}}(\bar{\theta})$ is calculated following Eq.~\eqref{eq:m_auto} or Eq.~\eqref{eq:m_cross} and depends on the angular-separation bin employed. 

\subsection{Shape noise}
\label{sec:noise}
Ignoring intrinsic alignment, intrinsic galaxy ellipticities and orientations are assumed to be uncorrelated and have a sampling distribution with mean zero \citep[e.g.][]{bernstein2002a}.
Its standard deviation, the intrinsic ellipticity dispersion $\sigma_{\epsilon}$ is also called shape noise.
We add this shape noise to our simulated maps, which will be introduced in Sect.~\ref{sec:simcomp}, as well as to the correlation-function likelihood calculation. 
For both the exact-likelihood part and the Gaussian covariance, we made use of the fact that white noise can be added to the full-sky power spectrum as a constant in $\ell$. 
The amplitude of the noise power spectrum $C_{\mathrm{n}}$ is determined by $\sigma_{\epsilon,1/2}$, the intrinsic galaxy shape dispersion per shear component ($\sigma_{\epsilon,1/2} = \sigma_{\epsilon} / \sqrt{2}$ ) and the area number density of source galaxies $n_{\mathrm{gal}}$:
\begin{equation}
\label{eq:noise_ps}
    C_{\mathrm{n}} =  \frac{\sigma_{\epsilon,1/2}^2}{n_{\mathrm{gal}}},
\end{equation}
where $n_{\mathrm{gal}}$ is given per steradian. 
The parameters $\sigma_{\epsilon,1/2}$ and $n_{\mathrm{gal}}$ do depend on the redshift bin in question but we assumed the same values of $\sigma_{\epsilon,1/2} = 0.28$ and $n_{\mathrm{gal}} = \SI{1.21}{\per \arcmin \squared}$ for both redshift bins under consideration.

This noise power spectrum can be propagated into the covariance of the correlation function, Eq.~\eqref{eq:gcov_xi}, by adding the noise power spectrum to each occurrence of the $E$- or $B$-mode theory power spectra \citep[e.g.][]{kilbinger2015a}.
We assume a constant galaxy density throughout but note that more realistic noise estimates are possible by taking into account the actually observed number of galaxy pairs \citep[for details, see for example appendix E of][]{joachimi2021a} or considering inhomogeneous noise \citep[][]{nicola2021a}.
For our exact likelihood, we also add the noise power spectrum, Eq.~\eqref{eq:noise_ps}, to the theory power spectra where applicable and let it propagate into the pseudo-$a_{\ell m}$ covariance matrix and therefore the exact correlation-function likelihood. 
All calculations leading to the full likelihood then follow exactly the same as in the noiseless case.
\subsection{Masks}
Throughout the presentation of the results, we will be employing three different masks. 
These include two simple circular cutouts, one with an area of $\SI{1000}{\sqd}$ and one with $\SI{10000}{\sqd}$, corresponding to roughly one fourth of the full sky.
One further, more realistic mask is taken from the KiDS-1000 survey \citep{joachimi2021a}, which also covers an area of about $\SI{1000}{\sqd}$.
All of the masks are created at or downsampled to a \texttt{HEALPix} $N_{\mathrm{side}}$ of $256$, corresponding to a pixelsize of $\ang{0.2}$ and implying a bandlimit of $\ell_{\mathrm{max}} = 767$.

Sharp edges of the mask in real space translate to ripples in harmonic space causing a ringing effect in the spherical-harmonic transform coefficients of the mask.
The normalization factor in Eq.~\eqref{eq:xip} contains these spherical-harmonic-transform coefficients of the mask. 
Likewise, the mask properties enter into the pseudo-${a}_{\ell m}$ mixing matrices $W_{\ell \ell' m m'}^{+/-}$ in their spherical-harmonic representation.

Since sums that involve the spherical-harmonic-transform coefficients of the mask might not converge when there is significant ringing present, the masks need to be smoothed. 
We applied a Gaussian smoothing filter that causes the mask ${a}_{\ell m}$ to drop to $10^{-6}$ of their original values at $\ell=30$.

We chose the three masks for demonstration purposes, but the method works for any mask, meaning it will perform just as well with larger footprints (as in stage-IV surveys) or more complex geometries.
However, the $\ell_{\mathrm{exact}}$ threshold for the exact calculation needs to be adjusted for every new mask, as we will show in Sect.~\ref{sec:ext_conv}.
\subsection{Simulated sampling distributions}
\label{sec:simcomp}
We compare our analytically predicted likelihoods to simulations by generating Gaussian random fields using the \texttt{HEALPix} function \texttt{synfast} from the same theory power spectra as used for the likelihood prediction.
Shape noise is added to the maps, where the noise level $\sigma_{\mathrm{n}}$ is determined by $\sigma_{\epsilon,1/2}$, the intrinsic galaxy shape dispersion as introduced in Sect.~\ref{sec:noise}, and the estimated number of galaxies per pixel given by the area number density $n_{\mathrm{gal}}$ times the pixel size $A_{\mathrm{pix}}$: 
\begin{equation}
    \sigma_{{\mathrm{n}},\mathrm{pix}} = \frac{\sigma_{\epsilon,1/2}}{\sqrt{n_{\mathrm{gal}} A_{\mathrm{pix}}}}.
\end{equation}
The pixel size $A_{\mathrm{pix}}$ corresponds to the \texttt{HEALPix} resolution $N_{\mathrm{side}} = 256$.

In principle, one could also add a constant noise term to the power spectra used to generate the signal maps, but it is a good check to see that the more natural approach of adding noise to each pixel after the map has been created results in the sampling distribution predicted by the likelihood where the noise is modeled via the power spectrum. 

The pseudo-$C_{\ell}$ are then measured using the \texttt{HEALPix} function \texttt{anafast}, after multiplying the generated fields with the same smoothed mask as used for the likelihood prediction. 
Finally, the obtained pseudo-$C_{\ell}$ are passed to the correlation function estimator, Eq.~\eqref{eq:xip}.
This approach makes it possible to measure the correlation function over a subset of the multipole moments in order to compare to exact likelihood calculations using only the $\ell_{\mathrm{exact}} + 1$ lowest multipole moments.

As a sanity check, we also run correlation function measurements on the same maps with a real-space correlation-function estimator using \texttt{TreeCorr}\footnote{\url{https://github.com/rmjarvis/TreeCorr}} \citep{jarvis2004a}.
We then compare these measurements to the pseudo-$C_{\ell}$ approach using the full range of multipole moments up to the bandlimit of the map (for details, see App.~\ref{app:corr_comp}).

To create correlated weak lensing maps for different redshift bins in order to measure cross-correlations, we use \texttt{GLASS}\footnote{\url{https://github.com/glass-dev/glass}} \citep{tessore2023a} with slight modifications to accommodate direct generation of correlated Gaussian shear maps from the corresponding auto- and cross-power spectra without having to go through the convergence. 
The procedures of adding noise and measuring the correlation functions follow as before.

In addition to the analytic Gaussian covariance introduced in Sect.~\ref{sec:full}, we will also compare our exact likelihood to a Gaussian likelihood with a covariance measured from these simulations. 
This gives us an estimate of a more sophisticated covariance, which takes the mask geometry into account. We will call this covariance the simulation-based covariance.

\section{Results}
\label{sec:res}
\subsection{One-dimensional likelihoods}
\begin{figure}
    \includegraphics[width=\columnwidth]{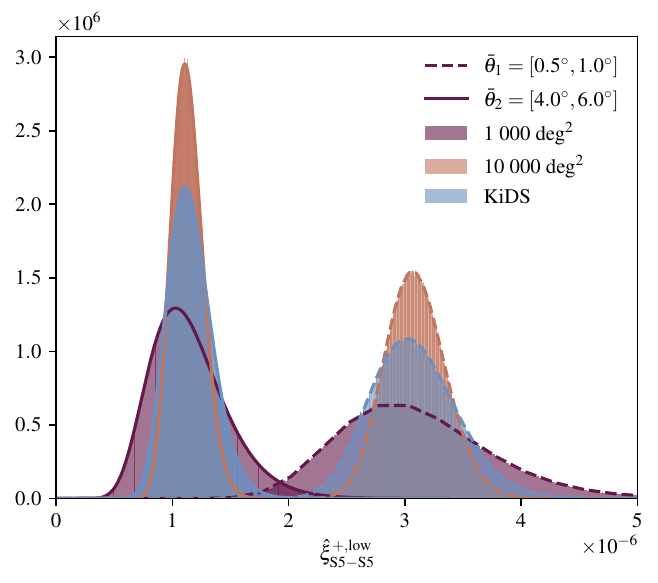}
    \vspace{-4mm}
    \caption{Histograms of simulated correlation functions (colored areas) and exact likelihoods for these correlation functions (dashed and solid lines) up to $\ell_{\mathrm{exact}}= 30$ (denoted by $\hat{{\xi}}^{{+, \mathrm{{low}}}}$) for different mask setups and two different angular-separation bins, $\bar{\theta}_1 = [ \ang{0.5}, \ang{1.0}]$ (dashed) and $\bar{\theta}_2 = [ \ang{4.0}, \ang{6.0}]$ (solid)}
    \label{fig:lowell}
\end{figure}
We begin our discussion of the results with the simplest one-dimensional case, where only one combination matrix and a one-dimensional characteristic-function-variable grid $t$ are required.
Following the theoretical considerations in Sect.~\ref{sec:ana}, we implemented a framework to calculate this one-dimensional likelihood of any weak-lensing correlation function given a theory power spectrum, a mask, and assuming the galaxy ellipticity field to be well approximated by a Gaussian random field (see App.~\ref{app:non_linear_scales}). 
\subsubsection{Exact likelihood}
\label{sec:exact_lh}
Considering only the exact low-multipole-moment part of the likelihood for now, denoted by $\hat{{\xi}}^{{+, \mathrm{{low}}}}$, we show the resulting likelihoods for a $\SI{1000}{\sqd}$, a $\SI{10000}{\sqd}$, and a KiDS-like mask in Fig.~\ref{fig:lowell} together with histograms of $10^6$ simulations with the same specifications and generated as described in Sect.~\ref{sec:simcomp}. 
We chose two angular-separation bins $\bar{\theta}_1 = [ \ang{0.5}, \ang{1.0}]$ and $\bar{\theta}_2 = [ \ang{4.0}, \ang{6.0}]$ to demonstrate that our framework works for different angular scales and to be able to compare the two, for example their degrees of skewness.
All distributions are shown for the case with added shape noise.
In the noiseless case, the skewness is even more pronounced and the distributions are narrower, as can be seen in Fig.~\ref{figapp:croco}. 
The predicted likelihoods perfectly match the simulated distributions and, as expected, the skewness tends to be more pronounced for the larger angular-separation bin and smaller mask area.
\subsubsection{Full likelihood and convergence}
\label{sec:ext_conv}
\begin{figure}
    \includegraphics[width=\columnwidth]{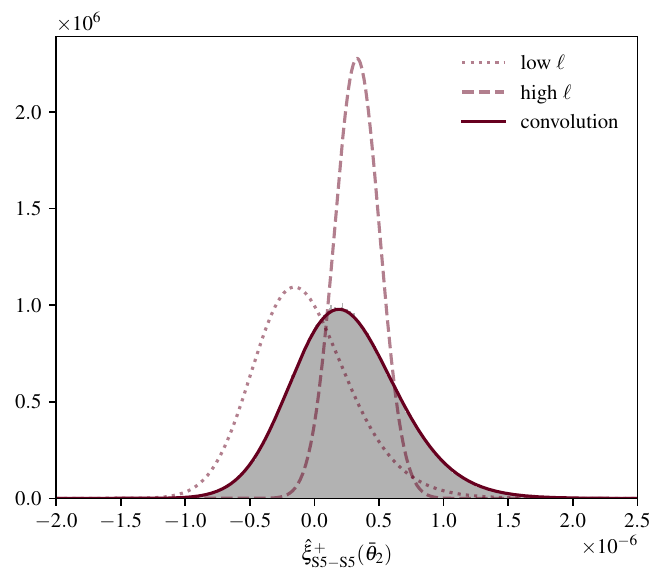}
    \vspace{-4mm}
    \caption{Histogram of simulated correlation functions up to the bandlimit of the map (gray area), exact low-multipole-moment likelihood (dotted), approximate high-multipole-moment likelihood (dashed) and the convolution of the two yielding the full likelihood (solid) for the $\SI{1000}{\sqd}$ mask and an angular-separation bin $\bar{\theta}_2 = [ \ang{4.0}, \ang{6.0}]$. The cutoff for splitting between exact and approximate part was chosen to be $\ell_{\mathrm{exact}} = 50$.}
    \label{fig:allell}
\end{figure}
Having established the likelihood up to a fixed and computationally feasible $\ell_{\mathrm{exact}}$, we now show the extension up to the bandlimit of a given survey.
In Fig.~\ref{fig:allell}, we use the example of the $\SI{1000}{\sqd}$ mask and the larger angular-separation bin, $\bar{\theta}_2 = [ \ang{4.0}, \ang{6.0}]$, and plot the (exact) low-multipole-moment part using an $\ell_{\mathrm{exact}}$ of $50$ (dotted) as well as the Gaussian approximation of the high-multipole-moment part (dashed) together with the convolution of the two yielding the approximation to the full $\ell$-range likelihood. 
We also show the corresponding simulations as gray histogram, again using $10^6$ realizations of the field.

For completeness, we show the one-dimensional likelihoods of a cross-correlation as well as the noise-free counterparts in App.~\ref{app:croco}.

We now briefly discuss the cutoff in the multipoles up to which the likelihood is calculated exactly and after which the Gaussian approximation is applied.
The mean, as already argued in Sect.~\ref{sec:full}, does not depend on this cutoff multipole moment, as the lower-$\ell$ mean is always given by the trace of $\mathbf{M}\boldsymbol{\Sigma}$, which is exactly the same as the (partial) sum over pseudo-$C_{\ell}$, and the high-$\ell$ mean is directly given by the other part of this sum.
The overall mean of a sum of two random variables is given by the sum of the individual means. 
The shape of the full $\ell$-range likelihood does depend on where this cutoff for the exact calculation is chosen, however. 
The question of how to choose it is related to the question from which scale onward the central limit theorem applies for a given mask and angular-separation bin. 
\begin{figure}
    \includegraphics[width=\columnwidth]{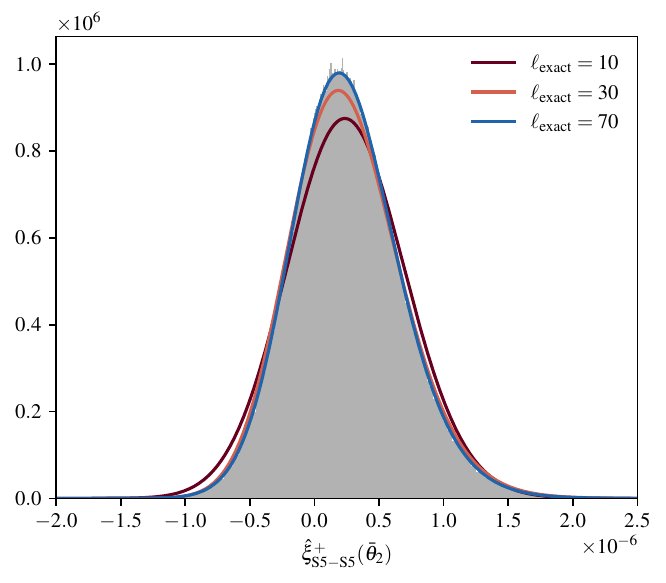}
    \vspace{-4mm}
    \caption{Histogram of simulated correlation functions up to the bandlimit of the map and the predicted full-$\ell$ range correlation-function likelihood for different values of $\ell_{\mathrm{exact}}$. This is shown for the $\SI{1000}{\sqd}$ mask and an angular-separation bin  $\bar{\theta}_2 = [ \ang{4.0}, \ang{6.0}]$.}
    \label{fig:ell_conv}
\end{figure}
To illustrate this, we show the full $\ell$-range likelihood for the same mask and angular-separation bin as before for different $\ell_{\mathrm{exact}}$ in Fig.~\ref{fig:ell_conv}.
The calculated likelihood approaches the simulated sampling distribution as the cutoff $\ell_{\mathrm{exact}}$ goes to smaller scales. 
This convergence can also be seen when looking at more quantitative descriptors of the distributions such as the variance or the skewness.
We show these quantitative results in App.~\ref{app:convergence} for the different masks.
As calculating the exact likelihood for the full range of multipole moments is computationally infeasible, we can only make an estimate of the cutoff based on these results.

We observe that this cutoff lies around $50$. 
This value will in practice depend on the setup, particularly, the chosen mask. 
Our results suggest that the maximum multipole moment needed for the exact calculation lies in a range that is feasible to achieve computationally for low-dimensional likelihoods as the one-dimensional example can still be calculated on a laptop. 
The $\ell_{\mathrm{exact}}$ needed could potentially further be lowered by using a more sophisticated estimate of the covariance for the (computationally cheap) high-$\ell$ part.

The observed level of agreement between the simulated distribution and the predicted likelihood shows that the Gaussian approximation with a simple $f_{\mathrm{sky}}$ covariance for the high-multipole-moment part of the sum and assumption of independence of the two parts can be justified.  
\subsection{Two-dimensional likelihoods}
\begin{figure*}
    \centering
    \includegraphics[width=\linewidth]{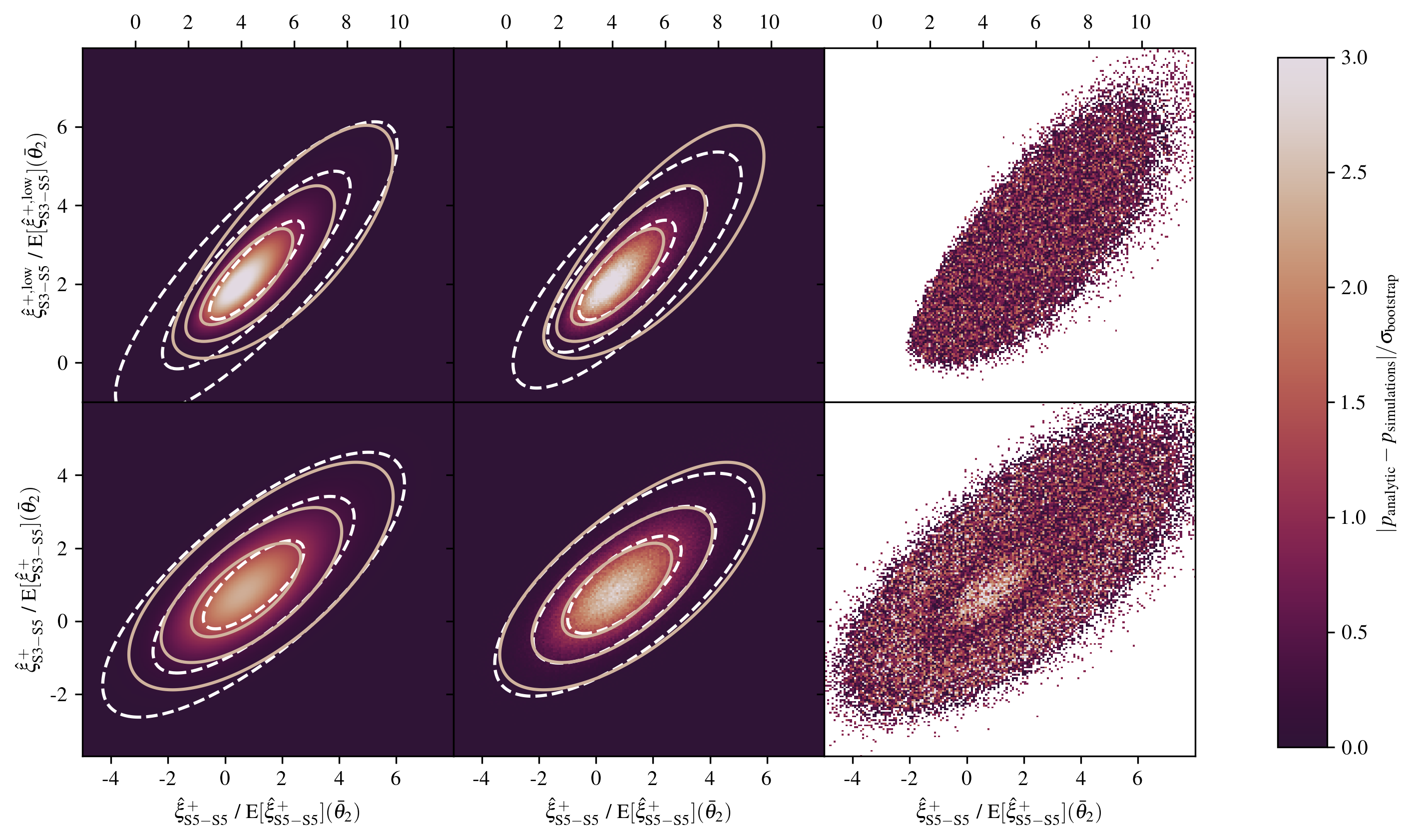}
    \caption{Two-dimensional joint likelihood of $\hat{\xi}^+_{\mathrm{{S5-S5}}}$ and $\hat{\xi}^+_{\mathrm{{S5-S3}}}$ for the angular-separation bin $\bar{\theta}_2 = [ \ang{4.0}, \ang{6.0}]$. In the left panel, the calculated exact likelihoods with corresponding contours (orange) and the Gaussian likelihood with analytic covariance (white dashed contours at $1$, $2$ and $3$ standard deviations) are shown. In the middle panel, histograms of the simulations are plotted together with the same contours, except that the Gaussian likelihood is shown with the simulation-based covariance. In the right panel, the residuals between analytic prediction and histogram of the simulated sampling distribution are displayed as multiples of the standard deviation of boostrap resamplings of the same histogram, which are quantified by the colorbar. The upper and lower row correspond to the (exact) low-multipole-moment part and the full $\ell$-range likelihood respectively. The domains are normalized to the mean of the full likelihood.}
    \label{fig:joint_lh}
\end{figure*}
To show that our framework can be extended to multidimensional likelihoods, we present an example of a two-dimensional correlation-function likelihood in this section.
In order to calculate the exact two-dimensional likelihood, two combination matrices $\bvec{M}^{\xi^{+}}_{ij} (\bar{\theta})$ and a two-dimensional grid of characteristic function variables $\mathbf{t}$ need to be set up.
We chose to show the joint likelihood for (cross-)correlations of two different redshift-bin combinations.
Namely, we arbitrarily chose the correlation functions $\xi^+_{S5-S5}$ and $\xi^+_{S3-S5}$ at the same angular-separation bin $\bar{\theta}_2 = [ \ang{4.0}, \ang{6.0}]$.
This could be changed to two different angular-separation bins at no additional effort, as only the argument of the corresponding combination matrix would need to be replaced. 
We show both the low-$\ell$ exact likelihood $\hat{{\xi}}^{{+, \mathrm{{low}}}}$ and the full $\ell$-range likelihood, i.e. the convolution of exact low-multipole-moment part and approximate high-multipole-moment part, together with their respective simulated distributions in Fig.~\ref{fig:joint_lh}.
The two-dimensional domain of the PDF is normalized to the mean of the respective full $\ell$-range likelihood $\operatorname{E}[\hat{\xi}^+_{ij}]$, meaning that the mean of the full $\ell$-range likelihood is always found at $\hat{\xi}^+_{ij} / \operatorname{E}[\hat{\xi}^+_{ij}] = 1$.
We overplot contours at $1$, $2$ and $3$ standard deviations of the corresponding multivariate Gaussian likelihood in white (dashed) and of the exact likelihoods at the same levels (orange) to highlight the difference in shape. 
The Gaussian likelihood is shown with the analytic covariance, Eq.~\eqref{eq:gcov_xi} with $\ell_{\mathrm{min}} = 2$, in the left column and with the simulation-based covariance in the middle column.
The contours mark the same levels.
We will discuss the comparison to the Gaussian approximation in the next section.

We show a more quantitative comparison between the analytic correlation-function likelihood and the simulated sampling distribution in the right column: to determine the sampling-noise level of the simulations we performed $500$ bootstrap resamplings of the simulated sets of correlated correlation functions.
Each of the resampled sets of simulations was then binned into the histogram just as for the simulations in the middle panel.
This way, we have a sampling distribution $p_b$ of the normalized distribution for each bin $b$.
The standard deviation $\sigma_b$ of each of these sampling distributions $p_b$ gives a good estimate of the scatter in each bin due to the finite number of simulations.
Finally, we consider the modulus of the residual between the normalized number of simulations from the original sample falling into each bin and the bin-wise predicted analytic likelihood $\vert p_{\mathrm{analytic}} - p_{\mathrm{simulations}}\vert$.
The right column of Fig.~\ref{fig:joint_lh} shows the fraction $\vert p_{\mathrm{analytic}} - p_{\mathrm{simulations}}\vert / \sigma_b$, i.e. how the residual compares to the estimated sampling-noise level, represented by the colorbar.

We have thus demonstrated that the difference between predicted and measured distribution is well below the noise level for the exact calculation and below the noise level except for a small fraction of bins for the full $\ell$-range calculation.
We note that the exact calculation has only been taken to $\ell_{\mathrm{exact}} = 30$ in the two-dimensional case for computational reasons.
For this particular angular-separation bin, mask, and $\ell_{\mathrm{exact}}$ the likelihood is not yet fully converged, as we have shown in Sect.~\ref{sec:ext_conv}.
Therefore, some discrepancy between the measured and predicted distributions is to be expected, particularly towards the mode and the tails of the distribution (see Fig.~\ref{fig:ell_conv},  $\ell_{\mathrm{exact}} = 30$ for the one-dimensional case), which is exactly what we see in the lower panel, right column of Fig.~\ref{fig:joint_lh}.

In principle, this approach can be extended to the full $n$-dimensional likelihood for all angular-separation bins and combinations of redshift bins used in a weak-lensing analysis.
Although the covariance matrix only needs to be calculated once per set of cosmological parameters, the eigenvalues need to be evaluated for each set $t_k$ of coordinates on the characteristic function grid for multidimensional distributions. 
Accounting for the scaling of the eigenvalue evaluation, the overall complexity is therefore $\mathcal{O}(N_{\mathrm{grid}}^{n} N_{\mathrm{field}}^3 \ell_{\mathrm{max}}^6)$ (see Sec.~\ref{sec:likelihood_charac}), where $N_{\mathrm{grid}}$ needs to be of $\mathcal{O}(10^3)$ to resolve the characteristic function.
Additionally, eigenvalue computations quickly become memory intense due to the storage of intermediate results.
Hence, this step becomes the computational bottleneck for multidimensional likelihoods, particularly when several redshift bins are considered where $N_{\mathrm{field}}$, and therefore the pseudo-$a_{\ell m}$ covariance matrix, becomes large.

\subsection{Comparison to the Gaussian likelihood}
\label{sec:gausscomp}
\begin{figure*}
    \centering
    \includegraphics[width=\linewidth]{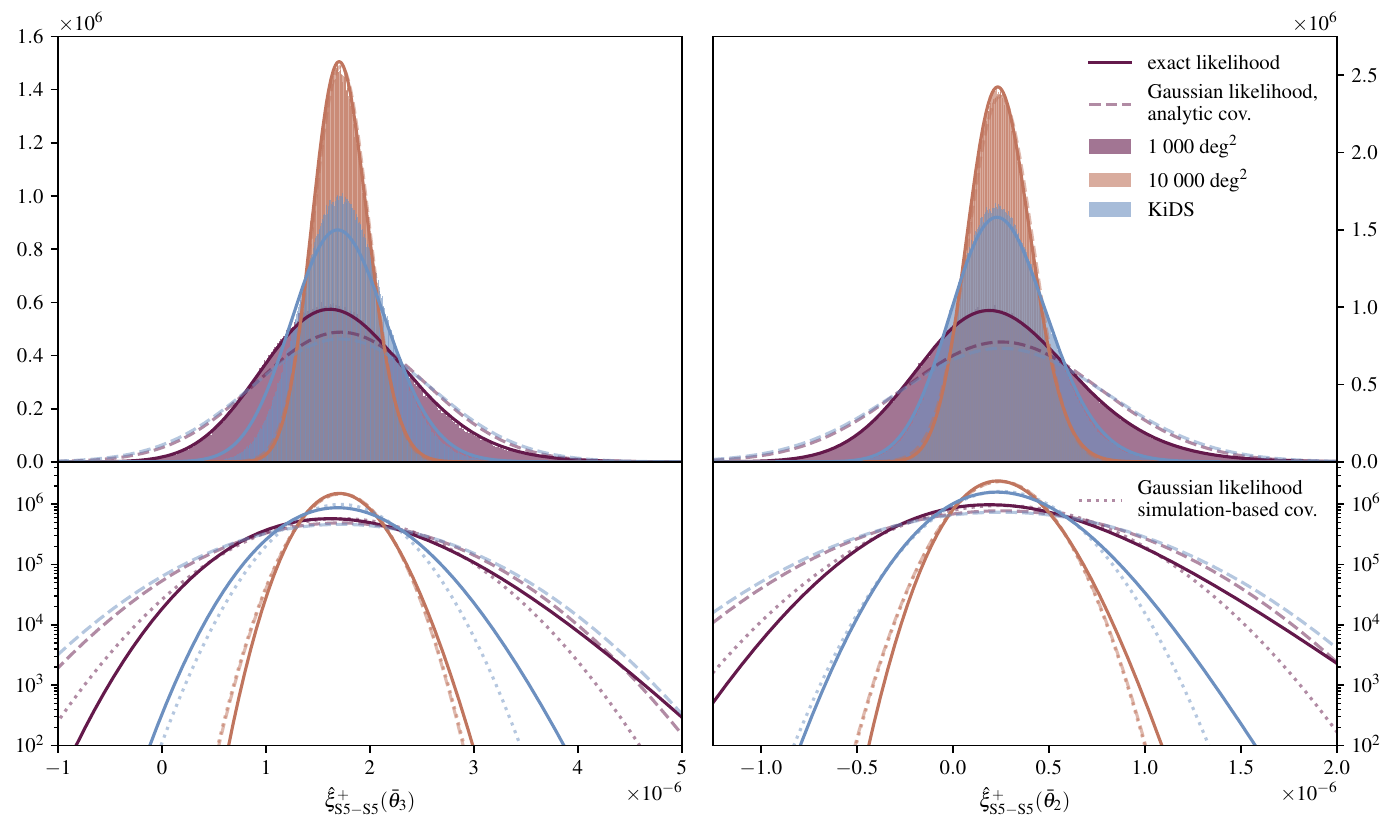}
    \caption{Full $\ell$-range exact correlation-function likelihoods (solid lines) for three different masks and two angular-separation bins: $\bar{\theta}_3 = [ \ang{2.0}, \ang{3.0}]$ (left column), and $\bar{\theta}_2 = [ \ang{4.0}, \ang{6.0}]$ (right column). Upper panel: Simulated correlation-function sampling distributions as histograms and the Gaussian likelihoods with analytic covariance (dashed lines). Lower panel: logarithmic presentation, the solid and dashed lines are the same as in the upper panel. Additionally, the Gaussian likelihoods are shown with simulation-based covariances (dotted lines).}
    \label{fig:gausscomp_1d}
\end{figure*}
\begin{figure}
    \centering
    \includegraphics[width=\linewidth]{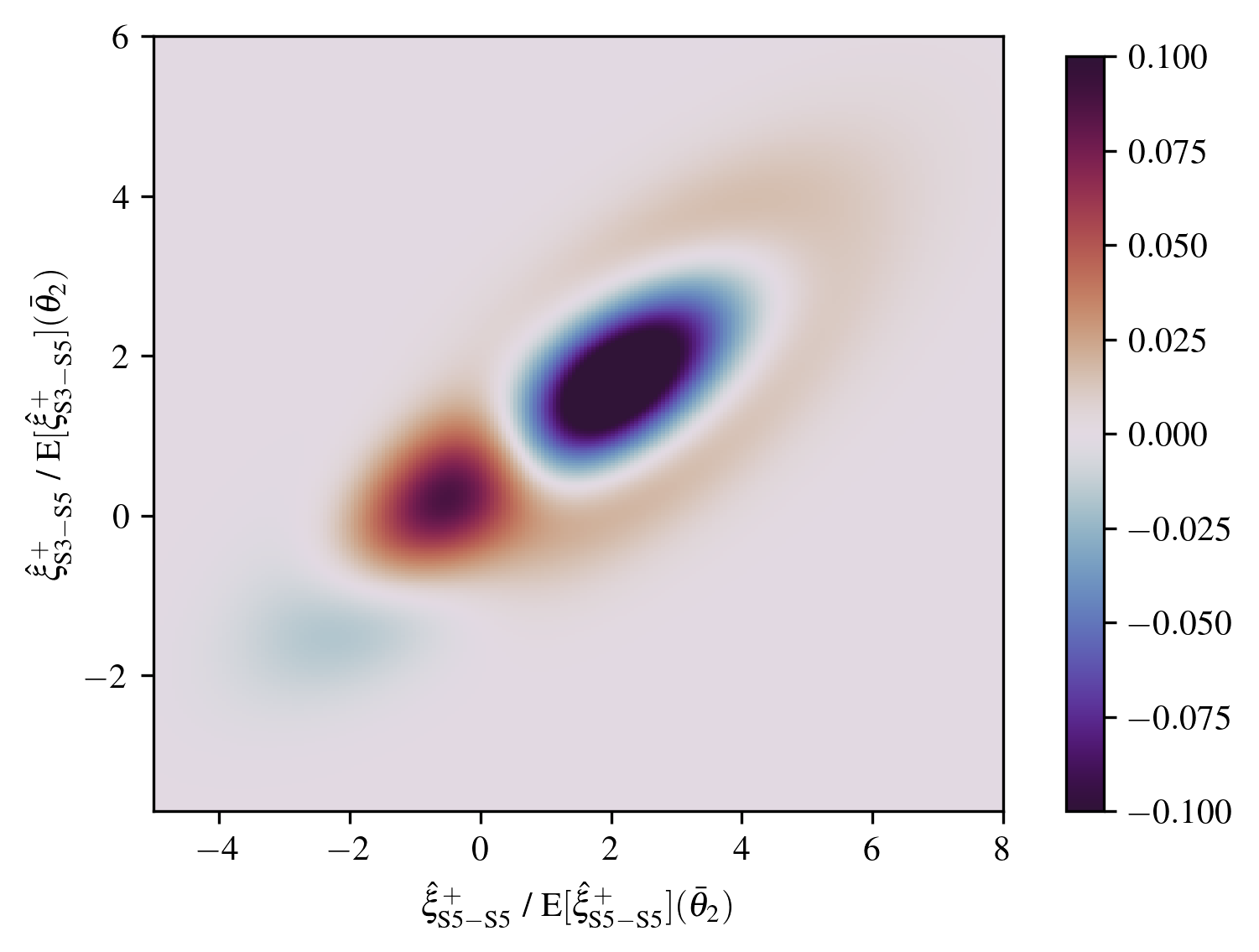}
    \caption{Comparison of the two-dimensional exact likelihood and Gaussian likelihood with the simulation-based covariance for the joint likelihood of $\hat{\xi}^+_{\mathrm{{S5-S5}}}$ and $\hat{\xi}^+_{\mathrm{{S5-S3}}}$ at the angular-separation bin $\bar{\theta}_2 = [ \ang{4.0}, \ang{6.0}]$. The setup is the same as in the lower middle panel of Fig.~\ref{fig:joint_lh} and the difference displayed with the colorbar is normalized to the maximum of the exact likelihood.}
    \label{fig:gausscomp_2d}
\end{figure}
To highlight the differences between our exact likelihood and the usually assumed Gaussian likelihood in a bit more detail, we directly compare the two likelihoods in this section.
Here, we will always mean the convolution of exact low-multipole-moment part and approximate high-multipole-moment part when referring to the ``exact likelihood'' in contrast to the fully Gaussian likelihood.
First, the Gaussian likelihood is obtained by using the mean and covariance as introduced in Sect.~\ref{sec:full} with the minimum- and maximum-multipole moment adjusted to the full $\ell$-range, i.e $\ell_{\mathrm{min}} = 2$ and $\ell_{\mathrm{max}} = 767$.
In Fig.~\ref{fig:gausscomp_1d}, we plot the full $\ell$-range exact likelihoods over the histograms of $10^6$ simulations for the three different masks (using $\ell_{\mathrm{exact}} = [50,70,100]$ according to the order in the plot) and two different angular-separation bins.
Note that the exact likelihood for the KiDS mask calculated with a cutoff $\ell_{\mathrm{exact}} = 100$ is not converged yet, which we also discuss in App.~\ref{app:convergence}.
For comparison, we plot the Gaussian likelihood as dashed lines.
We add a logarithmic presentation in the lower panel.
Here, the difference between exact likelihood and Gaussian likelihood becomes apparent, particularly in the tails of the distributions.

The difference between the exact likelihood and the Gaussian likelihood using the $f_{\mathrm{sky}}$ covariance approximation is quite substantial for the most complex mask, the KiDS mask, and is expected to be even larger when the exact likelihood is fully converged to the simulated sampling distribution.
For the KiDS mask and the $\SI{1000}{\sqd}$ mask, the Gaussian likelihoods with the $f_{\mathrm{sky}}$ covariance  are almost indistinguishable because they cover the same overall area but the exact calculation and the simulations show that the likelihood is in fact quite influenced by the particular mask shape.
To make the comparison fairer, we also retrieved a simulation-based covariance and plot the resulting Gaussian likelihood as dotted lines in the lower panel.

We now see the same trend for all masks and angular separations: the exact likelihood is higher towards higher correlation functions and a bit lower towards correlation functions left of the mode. The example of the KiDS mask with the smaller angular-separation bin is an exception but this might be due to the insufficient convergence of the exact likelihood using $\ell_{\mathrm{exact}} = 100$.
This trend is much less pronounced for the larger mask, where the difference between analytic and simulation-based covariance for the Gaussian likelihood is also barely visible.
We will use this $\SI{10000}{\sqd}$ mask, where the differences between exact and Gaussian likelihood seem smallest, in the next section to give an estimate of the impact on the posterior constraints.

Overall, the non-Gaussian exact likelihood cannot be reconciled with the Gaussian likelihood by choosing a better covariance for the Gaussian likelihood due to the apparent skewness of the exact likelihood.
So even with a perfect covariance estimate, a Gaussian likelihood will not be able to capture the complete distribution of correlation functions.
Interestingly, the skewness does not seem to change much for different angular-separation bins (see also Fig.~\ref{figapp:stats}) but it is stronger the smaller the mask area.

To illustrate the insufficiency of approximating the likelihood with a multivariate Gaussian even with a well-motivated covariance estimate, we plot the difference of the exact and Gaussian likelihood for the two-dimensional case normalized to the maximum of the exact likelihood in Fig.~\ref{fig:gausscomp_2d}. Positive values indicate that the exact likelihood is larger at a given point.
Even though the contours in Fig.~\ref{fig:joint_lh}, middle lower panel, seem quite close, suggesting very similar shapes, the picture changes when looking at the differences directly.
The structure of the exact likelihood is quite complex and deviates, particularly around the mean and mode, by more than $10 \%$ of the maximum of the exact likelihood from the Gaussian approximation.
The height of the mode exceeds that of the Gaussian mode and the distribution is wider in the tails towards higher correlation functions.
Meanwhile, it is lower at intermediate values of the correlation function just above the mean and for values below the mode.

We anticipate that higher-dimensional likelihoods, as needed for actual analyses, will increase in complexity.
Their behavior is difficult to predict without looking at simulated distributions or doing the exact calculation.

\subsection{Effect on posteriors}
\label{sec:post}
\begin{figure*}
    \centering
    \includegraphics[width=\linewidth]{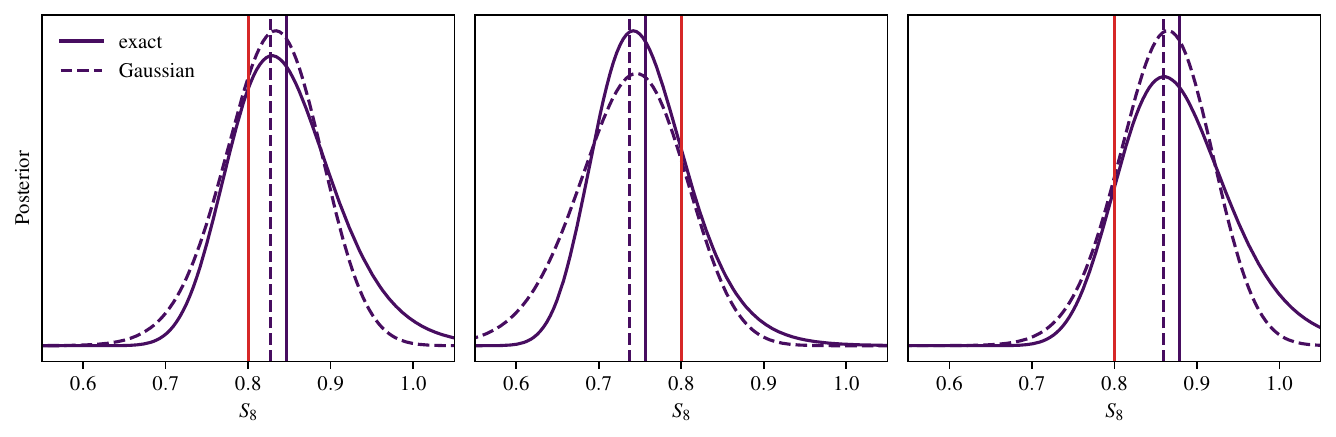}
	\caption{Three examples of posteriors obtained with our exact likelihood (solid) and the Gaussian likelihood (dashed) from different map realizations in comparison. The corresponding vertical lines mark the means of the respective distribution. We only varied one parameter and used a one-dimensional likelihood for the measurement of a correlation function value for exactly one angular-separation bin, $\bar{\theta}_3 = [ \ang{2.0}, \ang{3.0}]$. The $\SI{10000}{\sqd}$ mask was employed and a noiseless field was assumed. The red vertical line marks the value of $S_8$ used for the generation of the maps on which the measurement of $\xi^+_{S5-S5}(\bar{\theta}_3)$ was performed.}
	\label{fig:posts}
\end{figure*}
\begin{figure}
    \includegraphics[width=\columnwidth]{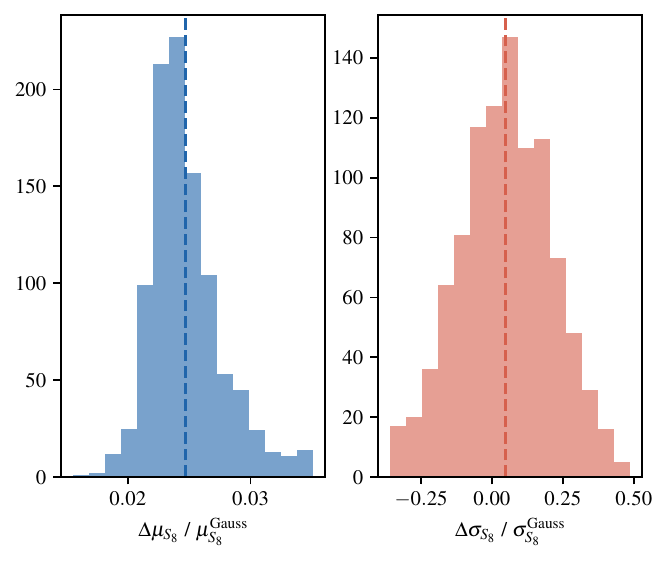}
    \vspace{-4mm}
    \caption{Histogram of the relative differences between posterior means and standard deviations obtained with the exact and Gaussian likelihood for $S_8$. The posteriors are calculated from $1000$ realizations of the $\xi^+_{S5-S5}(\bar{\theta}_3)$ measurement.}
    \label{fig:post_diffs}
\end{figure}
We describe the expected impact of using the exact likelihood instead of assuming a Gaussian likelihood on weak-lensing posteriors qualitatively at first:
especially the low-likelihood parts and tails of the exact likelihood (low meaning at $10\%$ or less of the maximum of the likelihoods when looking at Fig.~\ref{fig:gausscomp_1d}) deviate from the Gaussian likelihood.
That means that the likelihood will differ most when the tails of the likelihood are evaluated.
In general, higher values of the correlation function are more likely, lower values less likely as compared to the Gaussian approximation.
This is a trend we observe for all masks, given the exact likelihood is fully converged to the simulated sampling distribution.
The amplitude of the correlation function depends monotonically on the parameter $S_8$, which is a combination of the clustering parameter $\sigma_8$ and matter density parameter $\Omega_{\mathrm{m}}$, $S_8 = \sigma_8 \sqrt{\Omega_{\mathrm{m}} / 0.3}$, and can be constrained well by weak-lensing measurements, which is why it is also called the lensing amplitude.
One can therefore imagine, as higher values of the correlation function are preferred by the exact likelihood, that the posterior slightly moves towards higher $S_8$, in favor of further relaxing the $S_8$ tension. 

We can underpin this argumentation by running a small example.
For this example, we employed a one-dimensional likelihood for a measurement of $\hat{\xi}^+_{S5-S5} (\bar{\theta}_3)$ and varied only $S_8$.
We calculated $p(\hat{\xi}^+_{S5-S5} (\bar{\theta}_3) \vert S_8)$ exactly for different values of $S_8$ and set up the corresponding Gaussian likelihoods.
Assuming a Bayesian point of view on parameter inference and applying Bayes theorem, the two different posteriors of $S_8$ are then given as the product of these likelihoods and a prior on $S_8$.

We sample $S_8$ linearly between $0.4$ and $1.2$, which is equivalent to a uniform prior over this range and systematically explores the full prior range in a controlled setting. 
The posterior is directly given by the normalized likelihood evaluated at the measured value of $\hat{\xi}^+_{S5-S5} (\bar{\theta}_3)$ for each $S_8$.

We chose the $\SI{10000}{\sqd}$ mask to reduce sample variance and a noiseless realization of the lensing fields.
We also used the smaller angular-separation bin, $\bar{\theta}_3 = [ \ang{2.0}, \ang{3.0}]$, where the variation of the correlation function with $S_8$ is larger.
This is fairly restrictive but necessary because without these ideal conditions, one would not get any posterior constraint from a single angular-separation- and source-redshift bin.

For our exact likelihood, we chose a cutoff of $\ell_{\mathrm{exact}} = 30$, where we see a reasonable convergence for this mask and angular-separation bin (compare to Fig.~\ref{figapp:stats}).

The required theory $C_{\ell}$ are created by fixing $\Omega_{\mathrm{m}}$ and $\Omega_{\mathrm{b}}$ to $0.31$ and $0.046$ respectively and inferring $\sigma_8$ from the input $S_8$.
Otherwise assuming the same flat $\Lambda$CDM cosmology and source-redshift distributions as defined in Sect.~\ref{sec:wlapp} and picking the $S5$ bin for this application, these parameters are fed to \texttt{pyCCL} to retrieve the $S_8$-dependent sets of theory $C_{\ell}$.

For the Gaussian likelihood, we considered the analytic covariance here because for the $\SI{10000}{\sqd}$ mask, the analytic and simulation-based covariances are indistinguishable.
One thing to note here is that we compare to a Gaussian likelihood with fixed covariance, as this is the setup usually employed in weak-lensing analyses \citep[e.g. KiDS or DES;][]{joachimi2021a,friedrich2021a}.
Having a parameter-dependent covariance together with the Gaussian likelihood adds information such that posterior contours will become over-confident \citep{carron2013a}.
Hence, the best and closest to correct thing we can do is to use a constant covariance for the Gaussian likelihood and a naturally cosmology-dependent shape for the exact likelihood \citep[see also][]{wietersheim-kramsta2024a}. 

In Fig.~\ref{fig:posts}, we show a comparison of posteriors obtained with our exact likelihood and the corresponding Gaussian likelihood obtained from different realizations of the weak-lensing maps.
These examples show that the scatter of the posteriors $p(S_8 \vert \xi^+_{S5-S5} (\bar{\theta}_3))$ due to different realizations of $\hat{\xi}^+_{S5-S5} (\bar{\theta}_3)$ is larger than the differences due to the choice of likelihood.
The reason for this is twofold: one, at this relatively large angular separation, linear contributions to the power spectrum dominate the correlation function (see also App.~\ref{app:non_linear_scales}). The information content about clustering and therefore $S_8$ is smaller in these linear scales compared to the non-linear scales leading to less constraining power on $S_8$ in this part of the correlation function. 
Two, one angular-separation bin and one redshift-bin combination do not yield enough constraining power for a robust and consistently unbiased inference. 
Nevertheless, we find some characteristics that are common to all examples considered.

Across all retrieved examples of posteriors, we observe that higher values of $S_8$ are in fact slightly favored with our exact likelihood as compared to the posteriors obtained with the Gaussian likelihood.
The tail becomes more extended towards higher values of $S_8$ and flatter towards lower values of $S_8$ for the exact likelihood.
For a more quantitative result, we computed posteriors for $1000$ realizations of the data, always using the exact and Gaussian likelihood.
We recorded means $\mu_{S_8}^{\mathrm{Gauss / exact}}$ and standard deviations $\sigma_{S_8}^{\mathrm{Gauss / exact}}$ of these posteriors and display the relative differences $\Delta \mu_{S_8} / \mu_{S_8}^{\mathrm{Gauss}}$,  $\Delta \sigma_{S_8} / \sigma_{S_8}^{\mathrm{Gauss}}$, where $\Delta \mu_{S_8} = \mu_{S_8}^{\mathrm{exact}} - \mu_{S_8}^{\mathrm{Gauss}}$ and analogously for the standard deviation, between the exact- and Gaussian-likelihood cases in Fig.~\ref{fig:post_diffs}.
These histograms show that the mean is around $2.5 \%$ higher when obtained with the exact likelihood in a vast majority of cases, even though the means of the likelihoods are the same and a flat prior is assumed. 
We checked a few of the rather rare cases where the mean of the posterior obtained with the Gaussian likelihood is higher and noticed that these can be due to prior effects.
The deviation of the posterior mean by $2$ or $3 \%$ is an interesting observation because this difference roughly corresponds to the $68\%$ confidence interval given in current stage-III weak-lensing analyses for $S_8$ \citep[particularly, the joint analysis][]{dark-energy-survey-and-kilo-degree-survey-collaboration2023a}.
It is well possible that the difference in the posterior mean using the exact likelihood would be even larger for a smaller mask area as the differences between the exact and Gaussian likelihood are more pronounced. 
The distribution of standard deviations is rather symmetrical, such that posterior standard deviations seem to be comparable for the two different likelihoods employed, notwithstanding that the covariance of the likelihood varies with the cosmology in one case and not in the other.

Our example of one-dimensional posteriors suggests that a bias in the posterior mean can be caused when using a Gaussian likelihood, even with a fixed covariance. Meanwhile, the precision of parameters inferred using correlation functions does not seem to be influenced by the choice of a Gaussian likelihood instead of the exact one.

The impact of assuming a Gaussian likelihood instead of using the exact one on full multi-dimensional weak-lensing posteriors is difficult to capture at this point. 
This is because one would need the high-dimensional likelihood of all angular-separation bins and redshift-bin combinations for a given weak-lensing survey and be able to evaluate this likelihood for a large and also high-dimensional parameter space. 
With the current setup, likelihoods beyond a two- or three-dimensional one evaluated at a single point in parameter space are computationally very heavy due to the repeated retrieval of eigenvalues from large combination- and pseudo-$a_{\ell m}$-covariance matrices that themselves need to be recomputed for each point in parameter space.
As we have seen for the two-dimensional example, the likelihood becomes quite complex in shape when going to higher dimensions, such that its behavior and even more so the impact on the posteriors is at least difficult to extrapolate. 

Further investigations on how this could nevertheless be accomplished by speeding up the process or finding suitable approximations are left for future work.

\section{Conclusion}
In this work, we showed how the exact likelihood for correlation functions on masked spin-$2$ Gaussian random fields can be calculated. 
This also applies to joint likelihoods for in principle any number of dimensions.
The derived likelihoods match simulated sampling distributions of correlation functions very well. 
We applied our framework to a weak-lensing-survey setup and showed that the exact joint likelihood can be predicted up to the bandlimit of a given pixelated field by combining the exact low-multipole-moment likelihood with an approximate Gaussian likelihood for the small scales. 
Explicitly, we illustrated this with several one-dimensional and one two-dimensional example.
We compared our exact one-dimensional likelihood for several different masks and angular-separation bins to the widely used Gaussian likelihood and find that the differences are small when the sky coverage is large but the differing shape and a shifted mode could nevertheless lead to changes in the obtained posterior distribution for weak-lensing analyses. 
Particularly on large scales and even in the presence of shot noise we find that the deviation of the exact likelihood from the usually employed Gaussian likelihood can be substantial.
For stage-III weak-lensing surveys with survey areas comparable to the smaller masks we employ, these differences are more pronounced.

For the larger $\SI{10000}{\sqd}$ mask, where the difference between the exact likelihood and the Gaussian likelihood is small, we set up a comparison of $S_8$-posteriors obtained using our exact likelihood and the Gaussian likelihood. 
For the noise-free fields and a measurement of the auto-correlation function in one angular-separation bin, we see a change in the posterior mean using our exact likelihood as compared to the Gaussian likelihood. 
For different realizations of the correlation-function measurement, the posterior mean is found consistently at a higher value of $S_8$, meaning that the mean of this posterior is biased low by around $2.5 \%$ when using a Gaussian likelihood instead of the exact one. 
This does not allow for conclusions about the impact on the full posterior using noisy data, all angular-separation bins and redshift-bin combinations available yet, but strongly motivates further investigation, as data from stage-IV weak-lensing surveys is anticipated soon.

\citet{park2024a} show maximum likelihood results for $S_8$ inferred from many field realizations, similarly only varying $S_8$ and using noiseless fields but with the Gaussian likelihood. 
They use these results to show that they are consistent between using the spherical-harmonic and configuration-space two-point functions. 
But it is interesting that the distributions of retrieved $S_8$ are both in fact biased slightly low.

Looking ahead, the impact of replacing the approximate Gaussian likelihood with the exact one on the full posterior for upcoming stage-IV weak-lensing surveys needs to be evaluated.
An intermediate step would be to determine angular-separation ranges where the Gaussian likelihood is sufficient for specific survey setups.
Then, a mixed likelihood model could be employed, only adapting the exact likelihood for the largest scales.

Even for these largest scales, however, the computation of the exact likelihood has to be sped up.
We will investigate ways to achieve this in future work, for example by emulating the likelihood based on a smaller set of exact calculations or by first reducing the number of grid points for which the high-dimensional characteristic function is calculated.

Alternatively, analytic approximations to the likelihood can be considered once the exact form of the likelihood is known.

As we have already touched upon the calculation of the moments of our exact likelihood in Sect.~\ref{sec:likelihood_charac}, it is worth thinking about whether an Edgeworth expansion, as in \citet{lin2020a} but using cumulants as derived from the known moments instead of estimating them from simulations, would be a faster alternative.
This circumvents the need to calculate eigenvalues and reduces the problem to matrix multiplications and traces, which are easier to speed up.
It would still be required though to set up and calculate large covariance matrices for all pseudo-$a_{\ell m}$ needed for all correlation function estimators employed. 
As it is also possible that quite a few of the higher cumulants need to be calculated to approximate the exact distribution well enough and the Edgeworth expansion comes with other difficulties such as negative likelihoods, it remains to be seen whether this approach would simplify the exact part of the full likelihood. 
Another approximation worth considering could be a Gaussian mixture model which only requires the ability to sample from the distribution that should be parametrized. This is possible in the case of the correlation function but the high dimensionality of the problem is again a limiting factor and it is unclear whether the complex likelihood structure could be represented sufficiently. 

Simulation-based inference is a complementary approach to the exact calculations presented in this work.
The likelihoods learned in simulation-based inference could be compared to and validated against this exact likelihood on a subset of the parameter space, such that the exact likelihood does not need to be used in actual analyses but could be used as a benchmark and to estimate the reliability of learned likelihoods before they are used for higher-order or more complicated summary statistics. 
An example of posteriors obtained with a learned likelihood for two-point functions for a stage-III weak-lensing survey is given in \citet{jeffrey2024a} or \citet{wietersheim-kramsta2024a}.

In conclusion, the exact likelihood for correlation functions derived here can be used as starting point to model exact likelihoods for two-point functions in general. 
As we operate under the assumption that the fields are Gaussian random fields and the cosmological fields are not Gaussian on small scales, applicability of our results is limited to larger ($\gtrsim \ang{1}$) angular scales.
This is not necessarily a problem, as the likelihood for two-point functions naturally tends to become Gaussian on smaller scales.
Finally, our results are relevant not only for weak-lensing analyses but for all observables where two-point correlations are used including the CMB, \SI{21}{\centi \metre} mapping or measurements of the stochastic gravitational-wave background.

\section*{Acknowledgments}
\small
We acknowledge funding from the Swiss National Science Foundation under the Ambizione project PZ00P2\_193352.

\textbf{Software}: For this work, we made use of the Python packages \texttt{numpy} \citep{harris2020a}, \texttt{scipy} \citep{virtanen2020a}, \texttt{Wigner}\footnote{\url{https://github.com/ntessore/wigner}} \citep[based on][]{schulten1975a}, \texttt{pyCCL}, \texttt{GLASS}, \texttt{healpy} and figures were created with \texttt{Matplotlib} \citep{hunter2007a}.

\bibliographystyle{aasjournal}
\bibliography{library}
\newpage
\appendix
\section{On the assumption of Gaussian fields}
\label{app:non_linear_scales}
\begin{figure}
    \centering
    \includegraphics[width=0.5\columnwidth]{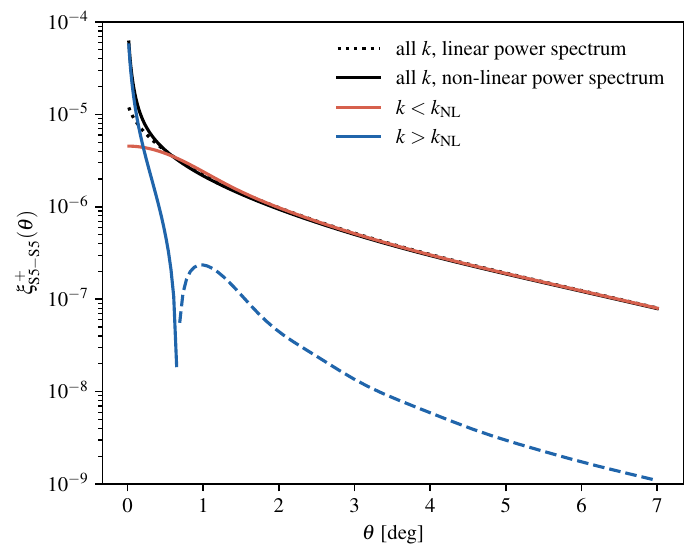}
    \caption{Correlation-function model prediction as a function of angular separation using the linear (black dotted) and non-linear three-dimensional power spectrum (black solid). The contributions from the large linear ($< k_{\mathrm{NL}}$) and smaller non-linear scales ($> k_{\mathrm{NL}}$) to the correlation function as derived from the non-linear power spectrum are shown in red and blue respectively. Negative (dashed line) and positive (solid lines) contributions are drawn separately in this logarithmic representation.}
    \phantomsection
    \label{figapp:k_contribs}
\end{figure}
In this work, we assume the cosmological fields to be Gaussian random fields.
To show that this assumption is valid, we estimate the contribution of non-linear scales to our correlation functions. 
The size scale $k_{\mathrm{NL}}$ in three-dimensional Fourier space marks the scale below which matter collapses non-linearly and is therefore the scale where the linear evolution of the primordial Gaussian matter density field breaks. 
At scales larger than $k_{\mathrm{NL}}$, meaning smaller scales in real space, the Gaussianity of the field and all derived or projected fields is therefore lost. 
This includes the weak-lensing fields we are concerned with. 
Using \texttt{pyCCL}, we split the three-dimensional matter power spectrum used to generate the two-dimensional projected weak-lensing power spectra into two regimes in $k$, that is above and below $k_{\mathrm{NL}}$. 
With this scale separation, we could consequently calculate the correlation function only using the linear or non-linear modes. 
As the comoving scale $k_{\mathrm{NL}}$ is redshift dependent but the three-dimensional power spectrum is evaluated along the line of sight, such that all redshifts from the source to the observer contribute to the weak-lensing power spectrum, we had to choose a redshift to select one $k_{\mathrm{NL}}$ on which we can base our scale cut.
We chose the redshift where our weak-lensing kernel peaks, i.e. where the three-dimensional power spectrum is weighted the strongest in the integration and therefore where the dominating contribution to the weak-lensing power spectrum is expected. 
This way, we find $k_{\mathrm{NL}} = \SI{0.14}{\mega \parsec \tothe {-1}}$.
We show in Fig.~\ref{figapp:k_contribs} that the contributions of the non-linear scales (blue lines), where the non-Gaussianity emerges, are subdominant by at least an order of magnitude and therefore do not contribute significantly to the large-angular-separation ($\gtrsim \ang{1.0}$) correlation function we exclusively consider in this work.
Therefore, the scales where non-Gaussianity is present in realistic cosmological fields do not enter the correlation-function estimates we use in this work.
The assumption of a Gaussian random field is therefore well valid for our considerations. 

\section{Definition of the mixing matrices}
\label{app:mixmat}
The mixing matrices $W_{\ell \ell' m m'}^{\pm}$ arise from the mask part $W(\bvec{\Omega})$ of the spin-weighted spherical-harmonic transform of the masked fields.
\begin{equation}
    \tensor[_{\pm 2}]{W}{}_{\ell \ell'}^{m m'} = \int \dd \bvec{\Omega} \ \tensor[_{\pm 2}]{Y}{_{\ell' m'}}(\bvec{\Omega}) W(\bvec{\Omega}) \tensor[_{\pm 2}]{Y}{}_{\ell m}^{\ast}(\bvec{\Omega}),
\end{equation}
where the $\tensor[_{\mp 2}]{Y}{_{\ell m}}(\bvec{\Omega})$ are the spin-$2$ spherical harmonics. 
For the pseudo-$a_{\ell m}$ equations, Eq.~\eqref{eq:palm}, one then defines
\begin{align}
\label{eq:wllmm}
    W_{\ell \ell' m m'}^+ &= \frac{1}{2} \left( \tensor[_{2}]{W}{}_{\ell \ell'}^{m m'} + \tensor[_{-2}]{W}{}_{\ell \ell'}^{m m'}\right)\\
    W_{\ell \ell' m m'}^- &= \frac{i}{2} \left( \tensor[_{2}]{W}{}_{\ell \ell'}^{m m'} - \tensor[_{-2}]{W}{}_{\ell \ell'}^{m m'}\right), \nonumber
\end{align}
see for example \citet{brown2005}. For the practical implementation for a discretized map, we will use the form given in \citet{Hamimeche_2009}, equation~B$15$, that only requires the spin-$0$ spherical-harmonic coefficients of the mask.

\section{Correlation function kernels}
\label{app:kernels}
\begin{figure}
    \centering
    \includegraphics[width=\linewidth]{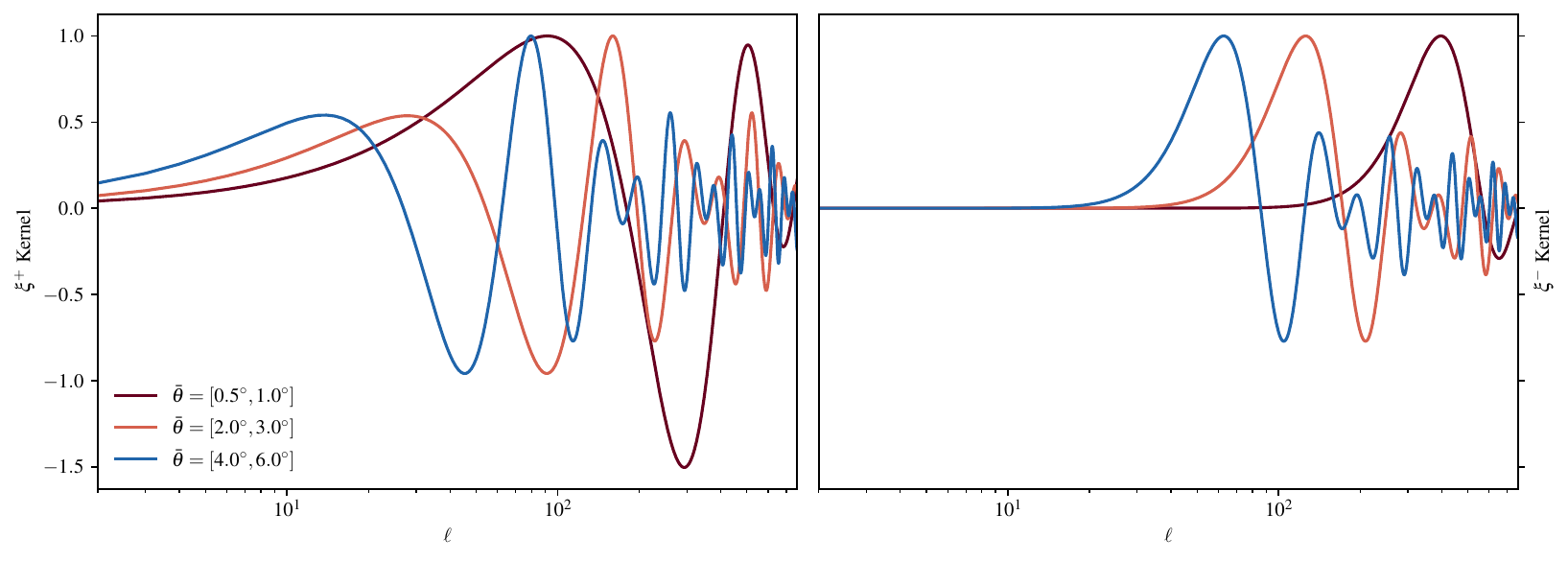}
    \caption{Kernels (i.e. summands) used in the transformation from pseudo-$C_{\ell}$ to the correlation function estimator Eq.~\eqref{eq:xip_cell} shown for both $\xi^+$ (left) and $\xi^-$ (right) for all angular-separation bins employed in this work. The functions are all normalized by their maximum value.}
    \phantomsection
    \label{figapp:xi_kernels}
\end{figure}
Large and small scales contribute differently to the correlation functions $\xi^+$ and $\xi^-$. This can be seen when looking at the weights for the pseudo-$C_{\ell}$ in Eq.~\eqref{eq:xip_cell}, which we call kernels in the following. 
The kernels for $\xi^+$ and $\xi^-$ read $(2 \ell + 1) K^+_{\ell} \left(\bar{\theta}\right)$ and $(2 \ell + 1) K^-_{\ell} \left(\bar{\theta}\right)$ respectively, where $K^+_{\ell}$ and $K^-_{\ell}$ summarize the angular dependence and bin integration, see Eq.~\eqref{eq:k_ell}, using the Wigner $d$-functions $d_{2 2}^{\ell}$ and $d_{2 -2}^{\ell}$ \citep[see also][]{chon2004}.

We demonstrate in Fig.~\ref{figapp:xi_kernels} that the $\xi^+$-kernel contributes more strongly at low $\ell$, whereas there is no significant contribution to $\xi^-$ below $\ell = 20$. These large scales, below $\ell = 20$, are the scales where the non-Gaussianity of the correlation-function likelihood mainly comes from. Therefore, and because the signal itself is stronger, we chose to present results only for $\xi^+$ in this work.

\section{Cross-correlation combination matrix}
\label{app:comb_mat}
For a cross-correlation of two different fields (e.g. across different redshift bins), the combination matrix needs to collect pseudo-${a}_{\ell m}$ of both fields and therefore looks like
\begin{equation}
    \label{eq:m_cross}
\bvec{M}^{\xi^{+}}_{ij} \left(\bar{\theta}\right) =
\begin{pmatrix}
    0 & 0 & \frac{1}{2} \bvec{M} (\bar{\theta}) & 0 \\
    0 & 0 & 0 & \frac{1}{2} \bvec{M} (\bar{\theta}) \\
    \frac{1}{2} \bvec{M} (\bar{\theta}) & 0 & 0 & 0 \\
    0 & \frac{1}{2} \bvec{M} (\bar{\theta}) & 0 & 0
\end{pmatrix}.
\end{equation}
In this case, the corresponding vector of pseudo-${a}_{\ell m}$ would be 
$\begin{pmatrix}
  \tilde{\bvec{a}}^{E,i} & \tilde{\bvec{a}}^{B,i} & \tilde{\bvec{a}}^{E,j} & \tilde{\bvec{a}}^{B,j}
\end{pmatrix}^T$,
where the bold $\tilde{\bvec{a}}^{X,i}$ denote vectors of all $\ell$ and $m$ for a given mode $X$, real and imaginary parts are both included, and they are sorted by $\ell$, then $m$.
The labels $i$ and $j$ distinguish different fields that could come from different redshift bins or even probes.
Each redshift bin or probe requires a set of pseudo-${a}_{\ell m}$, so the pseudo-${a}_{\ell m}$ vector will be twice as long, simply stacking these two sets.
The combination matrix is now designed that it does not just multiply the pseudo-${a}_{\ell m}$ with themselves but combines pseudo-${a}_{\ell m}$ from different redshift bins.
We note that the combination matrix has to be built in a way to bring the covariance parts, including the auto-correlations of both bins, into the characteristic function, meaning, that both permutations of the corresponding pseudo-${a}_{\ell m}$ are included and an additional factor $1/2$ is applied to the combination matrix to make up for the double counting. 
For a cross-correlation of two different redshift bins for example, cross-correlations can be calculated by using the corresponding theory cross-power spectrum. 
Since the pseudo-${a}_{\ell m}$ depend only on the ${a}_{\ell m}$ and the mask properties of their own field (albeit on the $E$- and $B$-modes of their own field), the covariance matrix can be set up block-wise for each set of corresponding full-sky auto- and cross-$C_{\ell}$. 

\section{Comparison between treecorr and the pseudo-\texorpdfstring{$C_{\ell}$}{Cl} correlation-function estimator}
\label{app:corr_comp}
\begin{figure}
    \centering
    \includegraphics[width=0.45\columnwidth]{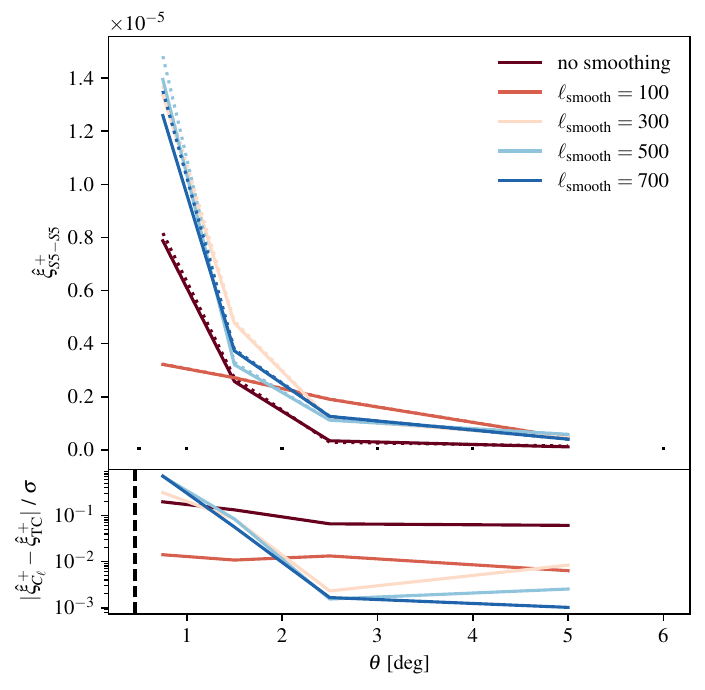}
    \caption{Comparison of correlation functions measured in the real space using \texttt{TreeCorr} (solid) and using the pseudo-$C_{\ell}$ correlation-function estimator (dotted) employed in this work. The underlying theory power spectra used to generate the maps are smoothed using a Gaussian cutoff with different $\ell_{\mathrm{smooth}}$, which is the multipole moment where the signal is suppressed to $10^{-6}$ of the original signal, shown in different colors. In this comparison, we used a circular mask with a $\SI{1000}{\sqd}$ cutout and an $N_{\mathrm{side}}$ of $256$. The angular-separation bin edges are shown as black dots. In the lower panel, we plot the absolute difference of the two estimators as a fraction of the standard deviation as given by the analytic Gaussian covariance. The dashed black vertical line marks twice the pixel size.}
    \phantomsection
    \label{figapp:sims}
\end{figure}
To show that our pseudo-$C_{\ell}$ correlation-function estimator, Eq.~\eqref{eq:xip_cell}, used to measure the correlation function on the masked Gaussian random fields recovers the correlation function in an unbiased way, we compared this estimate to a retrieval with the real-space estimator \texttt{TreeCorr}. 
With \texttt{TreeCorr}, we created a catalog by assigning coordinates and the corresponding value of the shear field to each pixel and then used the same mask as for the spherical-harmonic analysis by applying the corresponding weights. Fig.~\ref{figapp:sims} shows the correlation function for several applicable angular-separation bins measured with  \texttt{TreeCorr} (solid) and with our pseudo-$C_{\ell}$ correlation-function estimator (dotted). 
We applied different signal smoothings to see whether this has an impact on the agreement between the two. 
In fact, the agreement (shown in the lower panel as the absolute difference divided by the standard deviation as calculated from the covariance shown in Sect.~\ref{sec:full}) is better when the signal is smoothed quite strongly. 
Overall though, there is no clear trend, as the agreement seems to be better on small scales without any smoothing but better on the larger scales for any kind of smoothing, which is surprising as the signal is smoothed on small scales.
We also mark twice the pixel size (dashed vertical line) because pixelisation effects not taken into account in the signal power spectrum become important as one gets closer to the pixel size. 
Both the configuration-space measurement, where galaxy ellipticities are averaged to a pixel value for a map-based inference, and the spherical-harmonic analysis are impacted by this. 
For the measurement in spherical-harmonic space, the overall larger discrepancy on smaller scales can be explained by the fact that \texttt{HEALPix} spherical-harmonic analyses are only reliable up to an $\ell_{\mathrm{max}}$ of $2 N_{\mathrm{side}} = 512$ corresponding to an angular scale of $\ang{0.35}$. 
This issue does not impair our findings as we only consider correlation functions at the degree scale and it could be solved by going to higher map resolutions.
For our application, the agreement is given for any signal smoothing at a difference of $10 \%$ of the standard deviation or less such that we do not need to worry about signal smoothing or the correlation-function estimator we employ at this point. 

\section{Cross-correlation and noise-free likelihoods}
\label{app:croco}
\begin{figure}
    \centering
    \includegraphics[width=0.4\columnwidth]{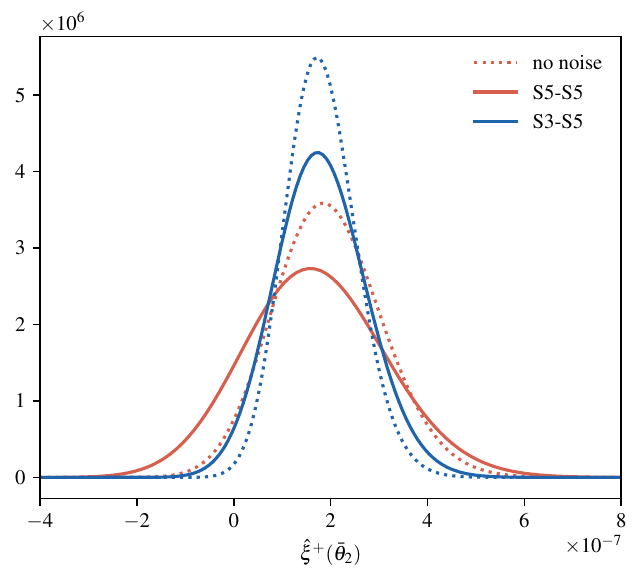}
    \caption{One-dimensional likelihoods comparing an auto- and cross-correlation function with (solid lines) and without (dotted lines) shot noise. The summation over multipoles is taken to the bandlimit of the map, i.e. includes the high-ell Gaussian extension, and we chose the $\SI{10000}{\sqd}$ circular mask and angular-separation bin $\bar{\theta}_2 = [ \ang{4.0}, \ang{6.0}]$.}
    \phantomsection
    \label{figapp:croco}
\end{figure}
Additionally to the one-dimensional analytical likelihoods for the auto-correlations shown in the main part of this work, we show likelihoods for an auto- and cross-correlation in comparison in Fig.~\ref{figapp:croco}, both evaluated at the angular-separation bin $\bar{\theta}_2 = [ \ang{4.0}, \ang{6.0}]$.
The cross correlation across redshift bins $S3$ and $S5$ exhibits less variance and the addition of noise does not increase the variance by a lot. For the auto-correlation case, adding noise clearly increases the variance and visibly decreases the skewness, i.e. pushes the likelihood towards a Gaussian distribution.

\section[Covergence in ell-exact]{Covergence in $\ell_{\mathrm{exact}}$}
\label{app:convergence}
\begin{figure*}
    \centering
    \includegraphics[width=\linewidth]{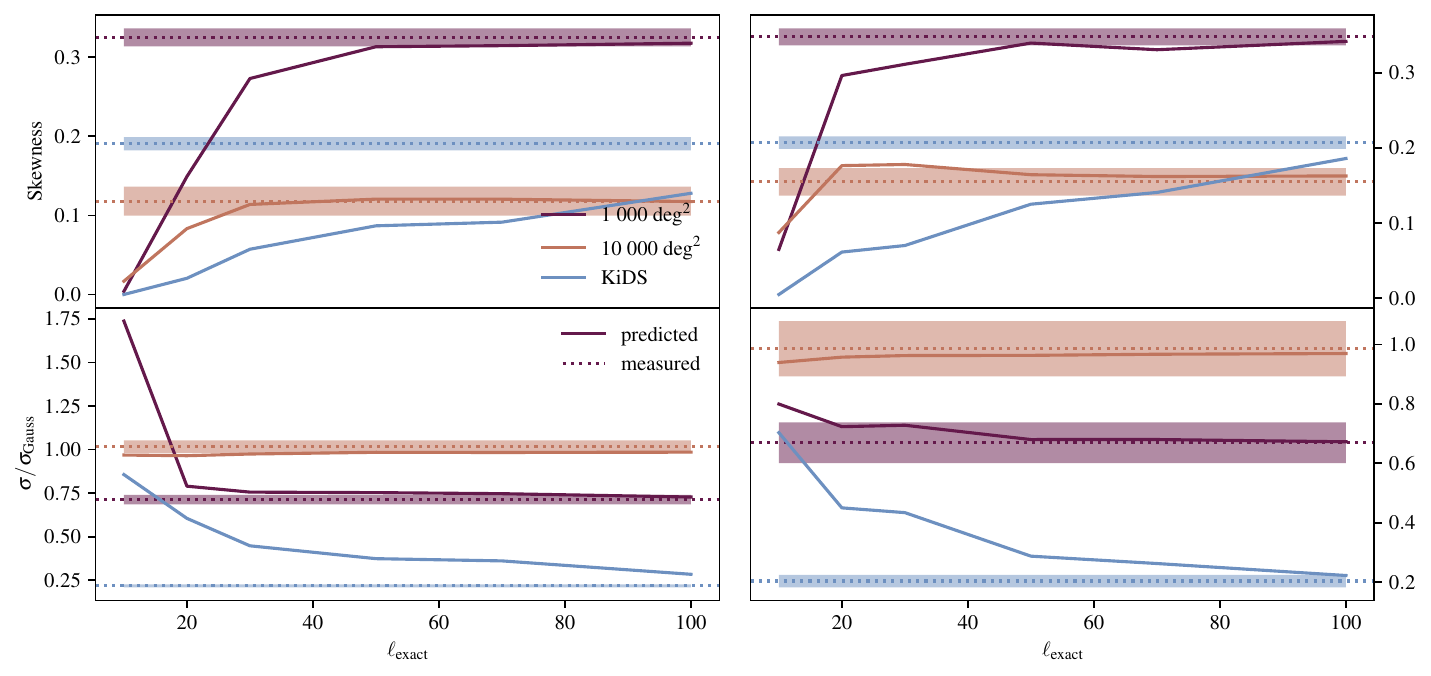}
    \caption{Skewnesses and standard deviations (displayed normalized to the corresponding Gaussian-likelihood standard deviation) of the exact likelihood for the three different masks and two angular-separation bins $\bar{\theta}_3 = [ \ang{2.0}, \ang{3.0}]$ (left) and $\bar{\theta}_2 = [ \ang{4.0}, \ang{6.0}]$ (right). The solid lines are the predictions from the exact-likelihood calculation as functions of the cutoff $\ell_{\mathrm{exact}}$. The horizontal dashed lines are the measured skewnesses and standard deviations from $300 \ 000$ map realizations. Standard deviations for skewness and standard deviation are estimated from $500$ bootstrap resamplings and shown as colored bands. }
    \phantomsection
    \label{figapp:stats}
\end{figure*}
To get an estimate up to which $\ell_{\mathrm{exact}}$ the exact low-$\ell$ likelihood needs to be calculated and from which onward the Gaussian high-$\ell$ extension can be applied, we calculated the full correlation-function likelihood for different cutoffs and compare the resulting skewnesses and standard deviations to each other but also to the simulated sampling distributions. 
In Fig.~\ref{figapp:stats}, we show the results for several $\ell_{\mathrm{exact}}$ between $10$ and $100$, all three example masks and two different angular-separation bins ($\bar{\theta}_3 = [ \ang{2.0}, \ang{3.0}]$ (left panel) and $\bar{\theta}_2 = [ \ang{4.0}, \ang{6.0}]$ (right panel)). 
The corresponding estimates on simulations have been performed on roughly $300 \ 000$ map simulations, where we performed $500$ bootstrap resamplings to estimate the standard deviation of the distribution characteristics displayed. 
These standard deviations are shown as colored bands. 
Convergence could be defined to be reached at the $\ell_{\mathrm{exact}}$ where the predicted value of all measures employed from the exact likelihood enters the $1 \sigma$ area of the ones measured from the simulations and does not change significantly anymore. 
Note that for the KiDS mask, convergence could not be reached yet with $\ell_{\mathrm{exact}} = 100$, particularly for the smaller angular-separation bin, hinting towards the complexity of the mask as an important factor determining the $\ell_{\mathrm{exact}}$ needed for the likelihood to pick up on complicated features. 
Overall, the skewness is slightly higher for the larger angular-separation bin and for a smaller mask area, as expected \citep[see e.g.][who show that the $n$-th moment scales as $1/f_{\mathrm{sky}}^{n-1}$]{lin2020a}.
Convergence could possibly be sped up by using a more sophisticated covariance estimate for the high-multipole-moment extension.

\end{document}